\documentclass[11pt]{article}
\usepackage{graphicx}
\usepackage[margin=1.25in]{geometry}
\usepackage[usenames,dvipsnames]{color}
\usepackage{url}
\usepackage{xcolor,hyperref,colortbl}
\usepackage{multicol}
\usepackage{comment}
\definecolor{lightgreen}{rgb}{0.53,0.66,0.42}
\definecolor{darkgreen}{rgb}{0.05, 0.5, 0.06}
\definecolor{lightblue}{rgb}{0.0,0.0,0.9}
\definecolor{orange}{rgb}{1,0.49,0.}
\definecolor{fucsia}{rgb}{1.0, 0.0, 0.5}
\definecolor{iceberg}{rgb}{0.44, 0.65, 0.82}
\definecolor{oldgold}{rgb}{0.81, 0.71, 0.23}
\definecolor{gray}{rgb}{0.66, 0.66, 0.66}

\usepackage{forloop}
\newcounter{loopcntr}

\newcommand{\on}[1][1]{
  \forloop{loopcntr}{0}{\value{loopcntr}<#1}{&\cellcolor{oldgold}}
}
\newcommand{\oni}[1][1]{
  \forloop{loopcntr}{0}{\value{loopcntr}<#1}{&\cellcolor{iceberg}}
}
\newcommand{\onx}[1][1]{
  \forloop{loopcntr}{0}{\value{loopcntr}<#1}{&\cellcolor{fucsia}}
}
\newcommand{\onb}[1][1]{
  \forloop{loopcntr}{0}{\value{loopcntr}<#1}{&\cellcolor{lightgreen}}
}
\newcommand{\ongg}[1][1]{
  \forloop{loopcntr}{0}{\value{loopcntr}<#1}{&\cellcolor{gray}}
}
\newcommand{\ong}[1][1]{
  \forloop{loopcntr}{0}{\value{loopcntr}<#1}{&\cellcolor{darkgreen}}
}
\newcommand{\onr}[1][1]{
  \forloop{loopcntr}{0}{\value{loopcntr}<#1}{&\cellcolor{orange}}
}
\newcommand{\off}[1][1]{
  \forloop{loopcntr}{0}{\value{loopcntr}<#1}{&}
}
\usepackage{tabularx}
 \usepackage{ltablex}

\newcommand\pubnumber{SLAC-PUB-17629}
\newcommand\pubdate{\today}

\def\Title#1{\begin{center} {\Large #1 } \end{center}}
\def\Author#1{\begin{center}{ \sc #1} \end{center}}
\def\corresponding#1{{ \textsuperscript{*} #1 }}
\def\Address#1{\begin{center}{ \it #1} \end{center}}

\newcommand\pubblock{\rightline{\begin{tabular}{l} \pubnumber\\
         \pubdate \end{tabular}}}
\newenvironment{Abstract}{\begin{quotation} \begin{center}
                       ABSTRACT
     \end{center}\bigskip  }{\end{quotation}}

\def\Acknowledgements{\bigskip  \bigskip \begin{center} \begin{large}
             \bf ACKNOWLEDGEMENTS \end{large}\end{center}}

\def\beq{\begin{equation}}
\def\eeq#1{\label{#1}\end{equation}}
\def\eeqn{\end{equation}}

\newenvironment{Eqnarray}%
   {\arraycolsep 0.14em\begin{eqnarray}}{\end{eqnarray}}
\def\beqa{\begin{Eqnarray}}
\def\eeqa#1{\label{#1}\end{Eqnarray}}
\def\eeqan{\end{Eqnarray}}

\let\bar=\overbar

\def\eg{{\it e.g.}}
\def\etc{{\it etc.}}

\def\lsim{\mathrel{\raise.3ex\hbox{$<$\kern-.75em\lower1ex\hbox{$\sim$}}}}
\def\gsim{\mathrel{\raise.3ex\hbox{$>$\kern-.75em\lower1ex\hbox{$\sim$}}}}

\def\del{\partial}
\def\Dslash{\not{\hbox{\kern-4pt $D$}}}
\def\dslash{\not{\hbox{\kern-2pt $\del$}}}

\def\Dlr{\mathrel{\raise1.5ex\hbox{$\leftrightarrow$\kern-1em\lower1.5ex\hbox{$D$}}}}

\def\ee{e^+e^-}
\def\sstw{\sin^2\theta_w}

\def\MSB{{\bar{M \kern -2pt S}}}
\def\msb{{\bar{\scriptsize M \kern -1pt S}}}

\def\drb{{\bar{\scriptsize D \kern -1pt R}}}

\makeatletter
\def\section{\@startsection{section}{0}{\z@}{5.5ex plus .5ex minus
 1.5ex}{2.3ex plus .2ex}{\large\bf}}
\def\subsection{\@startsection{subsection}{1}{\z@}{3.5ex plus .5ex minus
 1.5ex}{1.3ex plus .2ex}{\normalsize\bf}}
\def\subsubsection{\@startsection{subsubsection}{2}{\z@}{-3.5ex plus
-1ex minus  -.2ex}{2.3ex plus .2ex}{\normalsize\sl}}

\renewcommand{\@makecaption}[2]{%
   \vskip 10pt
   \setbox\@tempboxa\hbox{\small #1: #2}
   \ifdim \wd\@tempboxa >\hsize     
       \small #1: #2\par          
     \else                        
       \hbox to\hsize{\hfil\box\@tempboxa\hfil}
   \fi}

\makeatother

\def\CCC{C$^{3}$~}

\begin{document}
\begin{titlepage}
\pubblock

\vfill
\Title{\CCC:  A ``Cool'' Route to the Higgs Boson and Beyond}

\medskip 

\Author{Mei Bai, Tim Barklow, Rainer Bartoldus, Martin Breidenbach\textsuperscript{*}, \\Philippe Grenier, Zhirong Huang, Michael Kagan, Zenghai Li, \\ Thomas W. Markiewicz, Emilio A. Nanni\textsuperscript{*}, Mamdouh Nasr, Cho-Kuen Ng,\\ Marco Oriunno,  Michael E. Peskin\textsuperscript{*},  Thomas G. Rizzo, Ariel G.\\ Schwartzman, Dong Su, Sami Tantawi, Caterina Vernieri\textsuperscript{*}, Glen White, Charles C. Young}

\Address{SLAC National Accelerator Laboratory, Stanford University, Menlo Park, CA 94025}

\smallskip
\Author{John Lewellen, Evgenya Simakov}

\Address{Los Alamos National Laboratory, 
Los Alamos, NM 87545}

\Author{James Rosenzweig}

\Address{Department of Physics and Astronomy, University of California, Los Angeles, CA 90095}

\Author{Bruno Spataro}

\Address{INFN-LNF, Frascati, Rome 00044, Italy}

\Author{Vladimir Shiltsev}

\Address{Fermi National Accelerator Laboratory, Batavia IL 60510-5011}

\vfill

\begin{Abstract}

We present a proposal for a cold copper distributed coupling accelerator that can provide a rapid route to precision Higgs physics with a compact 8~km footprint.  This proposal is based on recent advances that increase the efficiency and operating gradient of a normal conducting accelerator.  This technology also provides an $\ee$ collider path to physics at multi-TeV energies.  In this article, we describe our vision for this technology and the near-term R\&D program needed to pursue it.

\end{Abstract}

\medskip

\corresponding{\footnotesize{mib@slac.stanford.edu, nanni@slac.stanford.edu, mpeskin@slac.stanford.edu, caterina@slac.stanford.edu}}
\vfill

\newpage

\tableofcontents
\end{titlepage}

\newpage

\def\thefootnote{\fnsymbol{footnote}}
\setcounter{footnote}{0}

\section{Introduction}

A top priority for the global particle physics community is to carry out precision studies of the Higgs boson using an $\ee$ collider.  In this article, we present a new proposal for such a collider based on recent advances in the technology of cold copper distributed coupling accelerators. Distributed coupling is a novel form of power distribution in the RF accelerator that allows for increased accelerator performance and better optimization. Cold copper refers to the operation of the accelerator at cryogenic temperatures to increase RF efficiency and the achievable accelerating gradient.
These new technologies are described by the acronym \CCC, or ``Cool
Copper Collider"~\cite{Bane:2018fzj, Nasr:2020acl}. Specifically, we propose a \CCC
linear collider on an 8 km footprint.  This collider can reach 250~GeV
in the center of mass energy (CM) using innovative technologies, with the possibility of a relatively inexpensive upgrade to 550~GeV on the same footprint.  This will achieve the goals of precision Higgs
boson and top quark measurements and it will at the same time 
provide a basis for the extension of $\ee$ physics into the multi-TeV energy range.

A facility of this type is urgently needed by the particle physics
community.  The urgency grows out of the great success of particle
physics over the past 30 years 
which, paradoxically, has led to increasing uncertainty about its future. We began
this period with tentative theories for the strong and weak sub-nuclear
interactions.  These theories were mathematically elegant, based on
Yang-Mills gauge theories, but stood on an uncertain experimental
footing.   Since 1989, we have confirmed the
predictions of these theories for the basic force carriers $W$ and $Z$
to the parts-per-mil level.  We have discovered the missing particles
predicted by this theory --- the top quark and the Higgs boson -- and
we have shown that their properties fit the theoretical predictions.
We have stress-tested the Cabibbo-Kobayashi-Maskawa model of flavor
mixing and CP violation and shown that so far it explains all 
experimental observations.  Finally, at the Large Hadron Collider (LHC), we
have observed complex reactions involving the strong and the weak
interactions and shown that the results  match the theoretical
predictions in all of their subtlety.  We can now confidently claim that the
``Standard Model'' of particle physics (SM) is established.

At the same time, we are more and more strongly persuaded that this
SM is incomplete.   It cannot account for the most
obvious properties of the universe at large --- the existence of the
dark matter and dark energy that make up 95\% of the mass of the universe and the 
large excess of matter over antimatter.  But even more troubling is
the fact that the SM is simply unable to answer the major questions
about its own structure:  Why do  its coupling constants take the
values that are observed?  Why are the masses of the quarks and
leptons so different, and what physics leads to the observed pattern?
Why does the Higgs field have a non-zero value in space, without which
there would be no mass spectrum for the other elementary particles?
It is now common to describe the SM as an ``effective'' theory that
should be derived from some more fundamental theory at higher
energies.  But we have almost no evidence on the properties of that
theory.

Our successes have become a liability in reaching this goal.  Scientists from other fields now have the impression that particle
physics is a finished subject.  They question our
motivations to go on to explore still higher energies.  This skepticism
confronts the reality that the next collider, whatever its technology,
 will be an expensive scientific instrument.  The scale of an
energy frontier collider is also challenging to the young people in
our field.   They need to see qualitatively new capabilities realized
during their active scientific careers. But it will require a concerted
effort to design and construct the next collider within 15 years from now. That is where the urgency lies.

To address this situation, our community needs to put forward an
ambitious program to discover the now-unknown fundamental interactions
that underlie the SM.   The community~\cite{icfa,efca} is converging globally on a program with
two essential elements:

\begin{enumerate}
\item  Construction of an $\ee$ Higgs factory:  Among all of the SM
  particles,
 the Higgs boson is the most
  central, the most tightly connected to the mysteries of the model
  and, at the same time, the particle least well studied.  With a
  next-generation $\ee$ collider, it will be possible to measure the
  properties of the Higgs boson to the 1\% level and below, a level
  that allows the discovery of effects from models beyond the SM.
  Several alternative technologies allow the construction of an $\ee$
  Higgs factory within 15 years, but this also requires a willing
  national host and a strong push from our global community.
\item  Development of multiple routes to an affordable  multi-TeV
  collider:  The first expectation for the energy scale of the new
  fundamental forces was the 1~TeV scale of Higgs boson physics.  From
  the LHC experiments, we now see that (barring some special scenarios) this physics is most likely to appear at a higher energy scale.  On the other hand, that scale cannot
  be orders of magnitude higher than 1~TeV if we expect to have a physics
  explanation for the Higgs field and its vacuum value.
  Reaching parton-parton CM energies of order 10~TeV should be the
  next  goal.   This goal is not realizable with any current
  technology.  Cost and power consumption are major issues.  We need
  to study as many feasible routes as possible, so that at least one route can succeed.
\end{enumerate}

For the first of these elements, the leading candidate now is the
International Linear Collider (ILC) in Japan~\cite{Bambade:2019fyw}.  The technical design is
virtually complete, and the technology is mature.  Still, there
remains the question of whether Japan will offer to host this project. On a longer time line, the Future Circular Collider (FCC)
electron-positron stage may provide a solution~\cite{FCC:2018evy}.  But this depends on
funding the needed 100~km tunnel, which costs as much as the LHC and
brings no physics capability in itself.   The Circular
Electron-Positron Collider (CEPC) in China may provide another
possible
solution~\cite{CEPCStudyGroup:2018rmc,CEPCStudyGroup:2018ghi}.
 As opposed to linear machines, circular colliders are strongly
 limited by 
synchrotron radiation above 350– 400 GeV center of mass 
energy.

For the second of these elements, ideas are being pursued based on
proton-proton colliders with high-field magnets~\cite{FCC:2018vvp},
muon 
colliders~\cite{muon1,muon2}, and
plasma-wakefield electron accelerators~\cite{ALEGRO:2019alc,pwawake}. The proton-proton route now
seems feasible but with a very high cost; the other routes have
technical barriers to overcome. 
 These very high energy options will require significant time
 and R\&D to realize.     

The development of cold copper distributed coupling accelerating cavities can
provide an 
alternative route to achieve both
steps in this program.   Our optimism about the capabilities of optimized copper accelerating cavities is based on new design ideas detailed in
\cite{Bane:2018fzj,Nasr:2020acl}.  The most important problem for operation of a normal-conducting cavity at high fields is electrical breakdown. 
Cavities  optimized for efficiency, high accelerating gradient and low breakdown have small irises that prevent power flow at the fundamental frequency.
 Individual feeds to each cavity from a common RF manifold, all in the same copper block, solve this problem.  Modern numerically controlled manufacturing techniques can build appropriate manifolds and individual cavity feeds in an extremely cost-effective way. We have also discovered that operation of these cavities at 80$^\circ$K increases their material strength and conductivity, giving marked improvements in performance. These two innovations lead to the  \textbf{\CCC
concept, 
a new elevated baseline for
normal-conducting electron accelerators}. 
%
%

Although there is no engineered and costed design for
a 250~GeV $\ee$ \CCC yet, this proposal is based on the SLC experience at SLAC and the extensive design work for ILC and CLIC.
  \CCC adds significant capabilities that allow robust designs with an
  accelerating gradient of 120~MeV/m.


This linac technology could be extended up to a 3 TeV collider by some combination of raising the gradient and extending the machine.    The primary challenges for the linac in the multi-TeV range are the cost and required power.   These can be mitigated by  highly efficient, low-cost RF power sources that are now being developed. Additionally, beyond 1~TeV beam-beam interactions increase the challenges associated with the beam delivery system and final focus. Beyond 3~TeV we enter \textit{terra incognita}. Here futuristic concepts such as muon colliders~\cite{muon1,muon2} or plasma wakefield accelerators~\cite{ALEGRO:2019alc,pwawake} may be required.  However, we believe that the \CCC  concept, augmented with additional new ideas, can also provide a route to these high energies.  We will discuss this issue in Sections \ref{sec:highergradient} and \ref{sec:beambeam}.


R\&D on the \CCC concept is already being pursued at SLAC, UCLA, INFN, LANL and Radiabeam, along with closely related research in high gradient RF acceleration with CERN, KEK, PSI, MIT, and many other partners in the high gradient research community~\cite{HG2021}.   There is 
direct synergy with the development of compact electron accelerators
for medical 
applications~\cite{Maxim2019,lu2021,Snively2020vhee,Nanni2020vhee} and the creation of compact X-ray free
electron lasers~\cite{rosen2020,graves2020} (FELs).  This technology will be further
developed, and it is expected to meet many of its initial goals within the Snowmass and
 P5 timelines, using our current
resources. The development of accelerators at the 100~GeV scale
and higher will require dedicated resources from HEP.

A \CCC $\ee$ collider could in principle be sited anywhere in the
world.  If the ILC goes forward in Japan, an energy upgrade using \CCC
accelerators could be built, re-using the ILC damping
rings, tunnel, and other conventional facilities.   However, it is important to note that 
the entire \CCC program could be sited in the
United States.  With the
cancellation of the Superconducting Super Collider and the end of
Tevatron operations the US
 has largely abandoned construction of domestic accelerators at the energy frontier. 
\CCC offers the opportunity to realize an affordable energy frontier
facility in the US.    This may be crucial to realize a Higgs factory
in the near term, and it will also position the US to lead the drive
to the next, higher energy stage of exploration.

The structure of this paper is the following:  In Section~\ref{sec:Higgs}, we will
sketch out a staged physics program leading to an $\ee$
collider for Higgs boson studies at 250 and 550~GeV.  Section~\ref{sec:acceler} will
present the near-term R\&D program for this collider, including work
that is already being pursued and a proposal for a \CCC demonstration
facility  reaching energies of a few GeV.  Section~\ref{sec:500} will describe the
design of the 550~GeV collider, explain the logic of the staged
upgrade path, and present a cost model.  Section~\ref{sec:TeV} will discuss how to
extend this technology to multi-TeV energies, explaining the issues
that must be solved and some possible pathways toward their solution.
Section~\ref{sec:Conclusions} gives our conclusions.

\section{Steps to a \CCC Higgs factory}\label{sec:Higgs}
 We describe a roadmap for \CCC to study the properties of the Higgs boson and improve other precision probes of the SM, and, eventually, to achieve energies that will allow the discovery and study of new particles beyond the reach of the LHC.   In this section, we will describe the evolution of a \CCC accelerator to its 550~GeV stage, which would achieve high-precision measurements of the couplings of the Higgs boson and the top quark.   The timeline that we propose for achieving this goal is shown in Table~\ref{tab:timeline}.


\begin{table}[h!]
\resizebox{\columnwidth}{!}{\noindent\begin{tabular}{p{0.22\textwidth}
!{\vrule width 0.5mm}p{0.01\textwidth}*{2}{|p{0.01\textwidth}}
!{\vrule width 0.3mm}p{0.01\textwidth}*{4}{|p{0.01\textwidth}}
!{\vrule width 0.3mm}p{0.01\textwidth}*{4}{|p{0.01\textwidth}}
!{\vrule width 0.3mm}p{0.01\textwidth}*{4}{|p{0.01\textwidth}}
!{\vrule width 0.3mm}p{0.01\textwidth}*{4}{|p{0.01\textwidth}}
!{\vrule width 0.3mm}p{0.01\textwidth}*{4}{|p{0.01\textwidth}}
|}
\hline
\textbf{}  & \multicolumn{3}{c!{\vrule width 0.3mm}}{\footnotesize{2019-2024}} 
           &\multicolumn{5}{c!{\vrule width 0.3mm}}{\footnotesize{2025-2034}} 
           & \multicolumn{5}{c!{\vrule width 0.3mm}}{\footnotesize{2035-2044}} 
           & \multicolumn{5}{c!{\vrule width 0.3mm}}{\footnotesize{2045-2054}} 
            & \multicolumn{5}{c!{\vrule width 0.3mm}}{\footnotesize{2055-2064}} \\
\hline
Accelerator \\
\hline
Demo proposal   \off[2] \oni[1] \off[20]  \\
Demo test   \off[3]\onx[3] \off[17]  \\
CDR preparation   \off[3]\oni[2] \off[18]  \\
TDR preparation   \off[5]\oni[2] \off[16]  \\
Industrialization   \off[6]\onx[2] \off[15]  \\
TDR review   \off[6]\oni[1] \off[16]  \\
Construction  \off[7] \onr[4]\off[12] \\
Commissioning  \off[10] \on[1]\off[5] \on[1]\off[6] \\
2 ab$^{-1}$ @ 250 GeV  \off[11] \onb[5]\off[7] \\
RF Upgrade   \off[11] \ongg[5]\off[7] \\
4 ab$^{-1}$  @ 550 GeV  \off[16] \ong[5] \off[2] \\
\footnotesize{Multi-TeV Upg.} \off[16] \ongg[7] \\
\hline
\hline
Detector \\
\hline
LOIs   \off[1] \oni[1] \off[21]  \\
TDR   \off[2]\onx[2] \off[19]  \\
Construction  \off[4] \onr[4]\off[15] \\
Commissioning  \off[8] \on[2]\off[13]  \\

\hline
\hline

\end{tabular}}
\caption{Timeline and Milestones for the proposed \CCC development}\label{tab:timeline}
\end{table}

The planned High Luminosity era of the LHC (HL-LHC), starting in 2027\footnote{This refers to the updated schedule presented in June 2021~\cite{LHCschedule}, as the LHC schedule is evolving, the starting date of HL-LHC could change.}, will extend the LHC dataset by a factor 10, and produce about 170 $\times 10^6$ Higgs bosons and 120,000 Higgs boson pairs. This will allow an increase in the precision for most of the Higgs boson couplings measurements~\cite{cepeda2019higgs}. HL-LHC will dramatically expand the physics reach for Higgs physics. Current projections are based on the Run 2 results and some basic assumptions that systematic uncertainties will scale with luminosity and that improved reconstruction and analysis techniques will be able to mitigate pileup effects. Studies based on the 3000 fb$^{-1}$ HL-LHC dataset estimate that we could achieve 2-4\% precision on the couplings to $W$, $Z$ and third generation fermions. But the couplings to first and second generation fermions will still largely not be accessible and the self-coupling will only be probed with O(50\%) precision. 

There are good reasons to measure the Higgs boson properties at higher precision than will be possible at the HL-LHC.   With the basic Higgs mechanism for mass generation now demonstrated, the next task for Higgs studies is to search for the influence of new interactions that can explain why the Higgs field has the properties required in the SM. If the new particles associated with these interactions are too heavy to be produced at the HL-LHC, they can still make an imprint on the pattern of Higgs boson couplings.  If we wish to prove the existence of these effects and to understand their pattern, we will need a very precise understanding of the Higgs boson properties, with measurements of 1\% or better.  Through global analyses, such high precision measurements can also improve the interpretation of LHC data and lead to a stronger comprehensive map of where new physics might lie. 

An $\ee$ Higgs factory has a distinctive duty to gain insight on the Higgs Yukawa couplings at the next level beyond the third generation fermions. In the SM, the Higgs Yukawa couplings are exactly proportional to mass.  This simplistic picture deserves close scrutiny. Tagging of charm and strange quarks, as previously demonstrated at SLC/LEP,  gives effective probes for advancing this program. There is a broader program to investigate the potential deeper connection of Higgs boson with flavor and CP violation. The cleaner $\ee$ environment aided by beam polarization could become a sensitive probe to reveal more subtle phenomena.

Studies for the four current $\ee$ Higgs factory proposals---ILC, CEPC, CLIC, and FCC-ee---demonstrate that experiments at these facilities can meet and even exceed these requirements for high precision.  Actually, despite their different strategies, the four proposals lead to very similar projected uncertainties on the Higgs boson couplings~\cite{deBlas:2019rxi}.  The improvement in the precision of Higgs boson couplings expected from the ILC in comparison to the HL-LHC is shown in Table~\ref{tab:smefthiggs}.  \CCC follows a program very similar to that proposed for the ILC.  We thus expect similar results, subject to considerations described below, for similar integrated luminosities and detector capabilities.  \CCC is expected to be less expensive for reaching the energy of 500-600~GeV.  But its key point is that it provides a more secure basis for extension to higher energies.

Although most of the gain in precision in Higgs boson couplings will be realized already at the 250~GeV stage, there are important reasons to continue the $\ee$ program to an energy of 500—600~GeV.   First, this energy is above the crossover point at which the $WW$ fusion reaction $\ee\to \nu\bar\nu h$ overtakes  the $\ee\to Zh$ reaction and becomes the dominant mode of Higgs boson production.   This means that, in  the 550~GeV data, the Higgs boson is mainly produced by a different mechanism with different sources of systematic errors.  A deviation from the SM predictions observed at 250 GeV can be cross-checked in this new dataset.  The  $\sigma\cdot BR$ for $WW$ fusion to $h$ with decay to $WW^*$ depends quartically on the Higgs-$W$ couplings and thus is a very powerful addition to the dataset. Second, this energy is needed to give a substantial cross section for the process $\ee\to Zhh$, which allows us to measure the Higgs self-coupling.The expected 20\% error on the Higgs self-coupling will allow us to exclude or demonstrate at 5~$\sigma$ the large enhancements to the self-coupling typically needed in models of electroweak baryogenesis~\cite{DiMicco:2019ngk}.   Third, this energy choice is well above the threshold for $\ee\to t\bar t$, far enough that top quark pairs are produced with relativistic velocities.  A short run
(200 fb$^{-1}$, about 6 months) at 350 GeV suffices to measure the {\it short-distance} top quark mass to a precision of better than 40~MeV. After this, a long run at 500-600~GeV would allow unambiguous separation of the effects of the various top quark electroweak form factors and measurement of these to the parts-per-mil level~\cite{Fujii:2015jha}.  

 While the CM energy of 250~GeV is very specifically called for as the peak of the $\ee\to Zh$ cross section, the optimal energy for the full machine is less precisely determined.  As explained above, the second stage of the \CCC should reach a high enough energy that the Higgs processes of  $WW$ fusion, $\ee\to \nu\bar\nu h$, and double Higgs bosons production $\ee\to Zhh$ have substantial cross sections and that top quarks should be produced through the process $\ee\to t\bar t$ at relativistic velocities. At 500-600~GeV,  it is also important to be far enough above the threshold for $\ee\to t \bar t h$ to provide sufficient statistics for the measurement of the top quark Yukawa coupling.   The statistical error on this measurement decreases by a factor 2 when one chooses 550~GeV rather than 500~GeV as the CM energy~\cite{Fujii:2015jha}.  In principle, higher energy gives more of an advantage, but this must be balanced against increased cost and footprint.  For the purpose of this paper, we have set the design energy of \CCC at 550~GeV.
 A crucial difference between models in which the Higgs boson is elementary and those in which the Higgs are composite is that, in the latter class, the top quark is also partially composite, with modified $W$ and $Z$ couplings. The 550~GeV measurements have great power to test for this property.   Moreover, at 550 GeV, an $\ee$ collider is as powerful as HL-LHC in searches for $Z’$ bosons and light quark and lepton compositeness and provides a complementary set of tests~\cite{fujii2019tests}.

Two differences between the \CCC and ILC accelerator designs should be noted.  First, our \CCC proposal assumes a conventional positron source with zero polarization, as opposed to 30\% for ILC. There is almost no difference between the two cases in the expected precision in Higgs boson couplings for a given luminosity~\cite{Fujii:2018mli}. Positron polarization does allow the collection of additional datasets that may be useful in controlling systematic errors.  \CCC is compatible with the addition of a polarized positron source as an upgrade.  For energies well above 500 GeV, the enhanced cross section for $e^-_Le^+_R$ collisions makes this advantageous. Second, the time structure of the electron and positron bunch trains is different, with \CCC having a bunch spacing of 3-5~ns as opposed to the ILC bunch spacing of 366~ns. 
While detailed simulation studies are required to fully assess the effects of the backgrounds in the detector, the existing ILC detector designs, SiD and ILD, would adapt well to \CCC. In particular, a dedicated optimization of the vertex detector design could absorb the different time structure without compromising the physics goals. Thus, we expect that the projected precision given in Table~\ref{tab:smefthiggs} for ILC will also apply to \CCC at 250~GeV.  It should be emphasized that the demands that the program of precision measurements places on systematic errors in tracking, heavy-quark tagging, jet energy measurement, and beam polarization measurement are very challenging for both machines.

%
%
While the acceleration technology of \CCC is well grounded, there are many technical aspects such as alignment and refrigeration flow that require a string test of at least three cryomodules, although simultaneous RF powering is not required.
This requires a demonstration facility for the \CCC technology.  Still, we foresee that the 250 GeV \CCC can begin operation close to or even before the end of data-taking at the HL-LHC as shown in Table~\ref{tab:timeline}.  
One should note that the luminosity of an optimized linear collider increases roughly proportional to the CM energy; thus, our timeline includes successively larger luminosity samples in equal intervals of time.

\begin{table}
\begin{center}
\begin{tabular}{|c | c | c | c |} 
 \hline
  Collider & HL-LHC & \CCC/ILC 250 GeV&  \CCC/ILC 500 GeV \\
  Luminosity   & 3 ab$^{-1}$  in 10 yrs  & 2 ab$^{-1}$   in 10 yrs   &  \textbf{+} 4 ab$^{-1}$   in 10 yrs   \\
  Polarization & - &  $\mathcal{P}_{e^+} = 30\%$ (0\%)&  $\mathcal{P}_{e^+} = 30\%$ (0\%)\\
 \hline\hline
$g_{HZZ}$ (\%)  & 3.2 & 0.38 (0.40) & 0.20 (0.21)\\
$g_{HWW}$ (\%) & 2.9 & 0.38 (0.40)& 0.20 (0.20) \\ 
$g_{Hbb}$ (\%) & 4.9 & 0.80 (0.85)  & 0.43 (0.44) \\ 
$g_{Hcc}$ (\%)  & - & 1.8  (1.8) & 1.1 (1.1) \\ 
$g_{Hgg}$ (\%) & 2.3 & 1.6 (1.7) & 0.92 (0.93)  \\
$g_{H\tau\tau}$ (\%) & 3.1 & 0.95 (1.0) & 0.64 (0.65) \\
$g_{H\mu\mu}$ (\%) & 3.1 & 4.0 (4.0) & 3.8 (3.8) \\
$g_{H\gamma\gamma}$ (\%) & 3.3 & 1.1 (1.1) & 0.97 (0.97)\\
$g_{HZ\gamma}$ (\%) & 11.  & 8.9 (8.9) & 6.5 (6.8) \\
$g_{Htt}$ (\%) & 3.5 & -- &  3.0 (3.0)$^*$ \\
$g_{HHH}$ (\%) & 50 & 49 (49)  &   22 (22) \\
$\Gamma_H$ (\%) & 5 &  1.3 (1.4) & 0.70 (0.70) \\ 
\hline
 \end{tabular}
\end{center}
\caption{ Precision on the Higgs boson couplings from SMEFT analysis (assuming flavor universality and no exotic decay modes) expected from HL-LHC and  ILC/\CCC~\cite{cepeda2019higgs,DiMicco:2019ngk,fujii2019tests}. All numbers are in \% and indicate 68\% C.L. probability to deviations in the different effective Higgs couplings. The $\ee$ 250 results include some LHC measurements, and the $\ee$ 500 results include the $\ee$ 250 data. The projections in the last column are computed for 500~GeV, following \cite{fujii2019tests}, except that for $g_{Htt}$ (marked with $^*$) the result given is for 550~GeV. Additional comparisons under different assumptions can be found in \cite{deBlas:2019rxi}. }
 \label{tab:smefthiggs}
\end{table}

The R\&D program to demonstrate the \CCC technology is described in Sec.~\ref{sec:acceler}.   The goal of this program is a ``string test'' of three cryomodules. In each of the 9~m cryomodules, 4 linac rafts (populated with two 1~m accelerating structures), two tunable cryogenic permanent quadrupoles and beam position monitors will be installed. The  cryomodules will demonstrate the full \CCC gradient of 120~MeV/m to accelerate an electron beam by 1~GeV per cryomodule. We believe this is the minimum required demonstration to propose the full accelerator.  We expect that the optimization of the cavity shape and the implementation of higher order mode suppression can be understood in the next three years, even before the construction of a demonstration facility.  The string test will then provide a verification of this 
program and a test of the stability of the system in continuous operation.   The demonstration accelerator may be directly useful for a driver of an X-ray free electron lasers, and, certainly,  5--10~GeV accelerators built with this technology will provide a new class of compact X-ray sources.

With the technology demonstration achieved, we will be ready to construct the accelerator that will be capable of reaching 500--600~GeV when powered to 120~MeV/m.  The other elements needed for a linear collider---the sources, damping rings, and beam delivery system---already have mature designs created for the ILC and CLIC.  Presently, our baseline uses these directly, however we will look for further cost optimizations for the specific needs of the \CCC. In Table~\ref{tab:LCparam} we make a direct comparison between the existing linear collider designs and \CCC.

\begin{table}[h!]
\begin{center}
\begin{tabular}{|c | c | c | c | c | c |} 
 \hline
  Collider & NLC\cite{nlc} & CLIC\cite{clic} & ILC\cite{Bambade:2019fyw} & \CCC & \CCC \\
   \hline\hline
   CM Energy [GeV] & 500 & 380 & 250 (500) & 250 & 550 \\
  $\sigma_z$ [$\mu$m] & 150  & 70 & 300   & 100  & 100  \\
  $\beta_x$ [mm]  & 10 & 8.0 & 8.0 & 12 & 12 \\ 
  $\beta_y$ [mm] & 0.2 & 0.1 & 0.41 & 0.12 & 0.12 \\
  $\epsilon_x$ [nm-rad]  & 4000 & 900 & 500 & 900 & 900\\
  $\epsilon_y$ [nm-rad]  & 110 & 20 & 35 & 20 & 20 \\
  Num. Bunches per Train  & 90 & 352 & 1312 & 133 & 75 \\
  Train Rep. Rate [Hz] & 180 & 50 & 5 & 120 & 120 \\
  Bunch Spacing [ns] & 1.4 & 0.5 & 369 & 5.26 &  3.5 \\
  Bunch Charge [nC] & 1.36 & 0.83 & 3.2 & 1 & 1 \\
  Beam Power [MW] & 5.5 & 2.8 & 2.63 & 2 & 2.45\\
  Crossing Angle [rad] & 0.020 & 0.0165 & 0.014 & 0.014 &  0.014\\
  Crab Angle & 0.020/2 & 0.0165/2 & 0.014/2 & 0.014/2 &  0.014/2  \\ 
  Luminosity [x10$^{34}$] & 0.6 & 1.5 & 1.35 & 1.3 & 2.4 \\
 & (w/ IP dil.) & (max is 4) & & & \\
Gradient [MeV/m] & 37 & 72 & 31.5 & 70 & 120 \\
Effective Gradient [MeV/m] & 29 & 57 & 21 & 63 & 108 \\   %
Shunt Impedance [M$\Omega$/m] & 98 & 95 & & 300 & 300 \\ %
Effective Shunt Impedance [M$\Omega$/m] & 50 & 39 & & 300 & 300 \\
Site Power [MW] & 121 & 168 & 125 & $\sim$150 & $\sim$175 \\ 
Length [km] & 23.8 & 11.4 & 20.5 (31) & 8 & 8 \\
L* [m] & 2 & 6 & 4.1 &  4.3 & 4.3 \\
 
\hline
 \end{tabular}
\end{center}
\caption{Beam parameters for various linear collider designs. Final focus parameters for \CCC are preliminary.}
 \label{tab:LCparam}
\end{table}

We believe that the most effective way to stage \CCC is to construct the complete set of cryomodules for 550~GeV together with an RF system that will produce an initial gradient of 70~MeV/m.   The level of RF power required is very close to that already available with modulators and klystrons that can be ordered from industry. 
Since the RF system is the primary cost driver for the Main Linac, this substantially lowers the cost of reaching 250~GeV.  In parallel with the design of the 250~GeV accelerator, we will pursue R\&D dedicated to dramatically lowering the cost of RF power and transferring these innovations to industry.   Then, when the 250~GeV machine is complete, we will be in a good position to purchase the additional RF needed for the upgrade to 550~GeV.

It may be possible to do physics at an intermediate stage in the construction at the $Z$ pole, 91 GeV. We do not consider this a part of our baseline, but we mention the possibility in case there is community interest. There is low-hanging fruit here. With a luminosity of $1\times 10^{33}$cm$^{-2}$s$^{-1}$, one-tenth of the 250 GeV design luminosity, a 2-year program would produce 1 billion polarized $Z$ bosons, enough to increase the precision on $\sstw$ and the polarization asymmetries $A_c$ and $A_b$ by more than an order of magnitude over the results of LEP and SLC.

This then describes the physics program for \CCC shown in Table~\ref{tab:timeline}.   Before 2040, and possibly coincident with the end of the HL-LHC run, the era of precision Higgs boson measurements would begin.   After 10 years of data-taking to achieve 2 ab$^{-1}$, half a million tagged Higgs bosons, the accelerator would turn on at the design energy of 550~GeV.  Here it would  acquire 4 ab$^{-1}$, giving an new, independent, dataset on the Higgs boson and a novel dataset on top quark production and decay.  Both the Higgs program and the top quark program will reach levels of precision at which one could realistically prove the presence of deviations from the SM. \CCC will also provide the first step toward realizing an accelerator at which the new particles responsible for those deviations could be produced and explored in detail.

\section{Current and near-future R\&D on \CCC acceleration}\label{sec:acceler}

The required development towards a 550 GeV \CCC is modest: The basic technology is extremely well demonstrated; the fundamentals of low emittance beams were established by SLC in the 1990's; and complete designs have been worked out for NLC, JLC and the ILC. Additional improvements for copper accelerators are based on high gradient research efforts and developments for CLIC and for low energy accelerator applications.



Very recently, there have been two key technological advances that have overcome previous limitations of this technology in reaching high field and low rates of electrical breakdown.  In this section, we will describe these advances and the path from our present understanding to a full machine proposal. We will first discuss the questions that are being addressed now in the \CCC R\&D program.   A complete machine design will require the successful operation of a \CCC demonstration facility.  We will describe the needed facility and its goals, and we will estimate the time scale and cost needed. 


\subsection{New design ideas for normal conducting accelerators}
\label{sec:newdesign}
\begin{figure}
\begin{center}
 \includegraphics[trim={0 5.5cm 6cm 6.5cm}, clip, width=0.95\hsize]{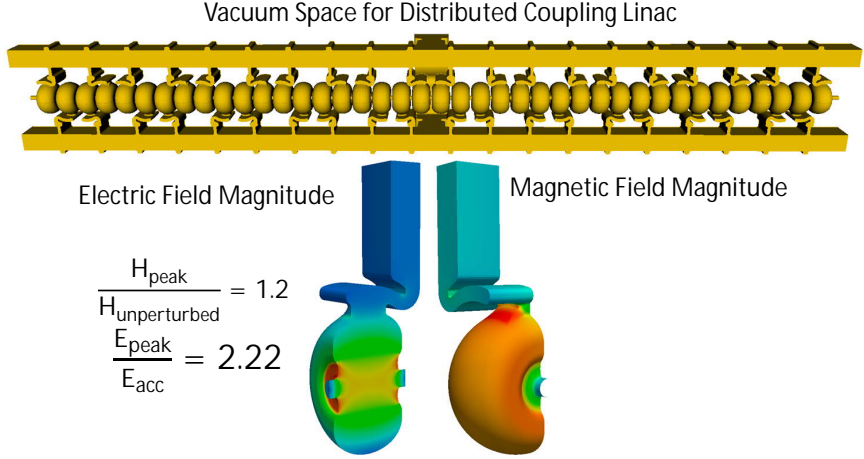} 
 \end{center}
\caption{(top) Vacuum space for a one meter long 40-cell C-band accelerating structure operating in the $\pi$ mode. (bottom) The magnitude of the electric and magnetic field in each cavity. Cavity geometries were optimized to limit the electric and magnetic fields strengths. The peak surface electric field to accelerating gradient ratio is 2.22. The perturbation to the magnetic field from the RF coupler increases the peak magnetic field by 1.2.}
\label{fig:DC}
\end{figure}

\begin{figure}
\begin{center}
 \includegraphics[trim={0 6.5cm 0 7cm}, clip, width=0.65\hsize]{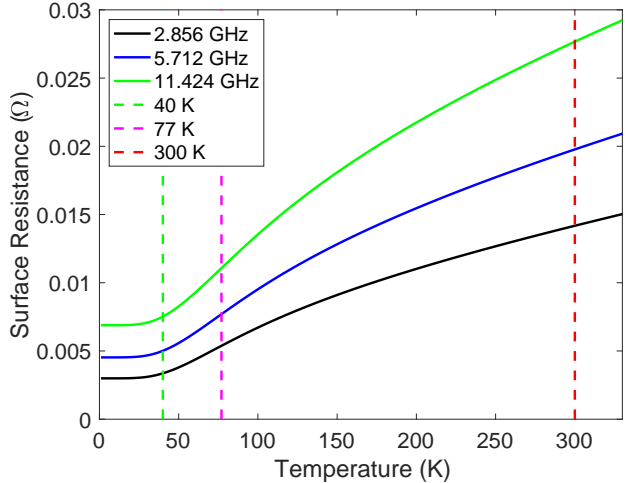} 
 \end{center}
\caption{Surface resistance of copper as a function of temperature including the anomalous skin effect. The reduction for surface resistance at 77K and 5.712 GHz (C-band) is a factor 2.55 compared to 300 K. \cite{Cahill:2016ipac,matula1979electrical}}
\label{fig:surfaces}
\end{figure}

The \CCC structure grew from a study of the fundamental limitations of high accelerating gradients and breakdown in normal conductors. Major improvements in fields and breakdown rates are possible with an optimized cavity shape that limits peak surface electric and magnetic fields, but with the seemingly problematic feature of an iris too small to propagate the fundamental RF mode. This led to the idea of an easily fabricated RF manifold and distributed coupling machined from the same blocks of metal as the accelerator. The distributed coupling powers the cavities at the correct phase and with equal fractions of RF power.  Developing this idea led to the two major advances:

\begin{enumerate}
\item RF power can be distributed to the cells of a copper accelerator individually through a distributed coupling waveguide.  With optimized cavity geometries, this scheme allows us to efficiently power the cells while maintaining strict limits on peak surface electric and magnetic fields.   A diagram of the structure's vacuum space is shown in Fig.~\ref{fig:DC}.   The structure can be mass-produced efficiently with currently used techniques.
\item Operating a copper structure at cryogenic temperatures allows us to further reduce RF power requirements while increasing the achievable beam loading and accelerating gradient, through a combination of increased material strength and reduced strain.  A key parameter, the surface resistance, is shown as a function of temperature in Fig.~\ref{fig:surfaces}.  Almost all of the improvement is achieved by operation at liquid nitrogen temperature, 77$^\circ$K. 
\end{enumerate}

The result of these two advances is a reduction in the required RF power to establish the gradient by greater than a factor of six compared to other normal conducting accelerators with similar beam aperture, see Table~\ref{tab:LCparam}. The RF to beam power efficiency increases significantly to 45$\%$ and plays an important role in the feasibility of extending \CCC to the multi-TeV scale. The achievable accelerating gradient also increases.  The  full collider design we propose here has a gradient of 120~MeV/m, but this number is not limited  by the accelerator technology, but rather by an overall optimization with the cost of  RF power.  As we will discuss in Sec.~\ref{sec:TeV}, engineering improvements here can provide a path to much higher energies.

RF accelerators can operate over a wide range of frequencies. Most RF accelerators developed for a linear collider operate between 1-12 GHz. A detailed study of cavity geometry for 1~nC bunches shows that C-band (5.712~GHz) is the optimal frequency to realize \CCC.  The principal trade-offs are a reduced structure efficiency at lower frequency, but better beam dynamics for the 1~nC electron bunch train. C-band allows us to maintain a highly efficient structure, high gradient operation and excellent beam dynamics with proper damping and detuning of the cavities.

\CCC utilizes split-block fabrication to produce $\sim$1~meter structures with all cavities machined out of two or four copper slabs. The cavities are machined with modern CNC methods and the cavity surfaces require no post CNC machining,  greatly reducing the production cost.  Bonding of the two halves is needed, but the bonding is parallel to the direction of current flow and has an extremely small impact on the RF performance of the accelerating mode. Higher-order mode detuning can be  incorporated by adjusting the cavity geometry of each cell during fabrication.  This is already a standard procedure. Slot damping with lossy materials will suppress long range wakefields further.  

These two ideas are the basis for a new, highly efficient generation of copper RF accelerators.  We now turn to the question of how these features can be demonstrated to give a firm foundation for the \CCC design.

\subsection{Optimization of the RF accelerating structure}

The final accelerator structure for \CCC including wakefield damping and detuning will be fabricated and tested in the first year of the demonstration facility described in the next section. This meter scale structure will integrate all of the advanced concepts and meet all of the technical requirements for \CCC, including operation at high gradient. However, many of these key concepts and design elements have already been demonstrated using current laboratory facilities, and we continue to advance the state of the art to increase the technical readiness of \CCC.

High gradient operation of a distributed coupling accelerator at normal and cryogenic temperatures has already been demonstrated  (including acceleration of beam) in multiple single and multi-cell structures~\cite{Bane:2018fzj,Nasr:2020acl,cahill2018,tantawi2020}.   High gradient operation with low breakdown rate has been demonstrated at C-band~\cite{simakov2021}. A one-meter \CCC prototype was fabricated using the desired split-block technique and passed low power RF tests at cryogenic temperatures~\cite{nanni2021}.  Slot damping features were incorporated into a prototype structure without reducing RF performance. Cryogenic RF properties were measured at C-band,  demonstrating a reduction of a factor of more than 2.5 in RF surface resistance~\cite{nannihg2021}. RF designs for cavity detuning and cavity damping to meet the requirements for a multi-TeV collider have been developed.  Additional structures are now being produced and tested.  Since testing does not necessarily require cryogenic operation, we expect that the structures can be quickly optimized.

Distributed coupling also allows one to optimize the phase advance between cells while adjusting for the cavity length. For the beam aperture in our design, a cavity phase advance of 135$^\circ$ is optimal and increases the shunt impedance (133 M$\Omega$/m) by nearly 10$\%$ over the $\pi$-mode (120 M$\Omega$/m). The structure design assumed in this paper  is a $\pi$-mode accelerator with a factor 2.5 improvement in shunt impedance at 77$^\circ$K,  as was measured in our prototype. However, we will continue to improve this design up through the initial phase of the demonstration facility, in so far as it does not impact the cryomodule design. This approach provides a 
300~M$\Omega$/m shunt impedance for its planned mode of operation (beam aperture radius 2.62~mm). For comparison, NLC structures operated with a shunt impedance of up to 98~M$\Omega$/m for the periodic cells (with a standing wave equivalent shunt impedance of 50~M$\Omega$/m)~\cite{nlc},
albeit with a larger aperture (beam aperture radius 3.75~mm).
CLIC structures currently under consideration operate with a shunt impedance of 95.4~M$\Omega$/m (with a standing wave equivalent shunt impedance of 39~M$\Omega$/m) and a beam aperture radius 3.15-2.35 mm~\cite{clic}.

Combining all of these advances into a single mass-producible accelerating structure remains a high priority that must be completed early in the demonstration facility phase of the program.  We plan that the complete damped and detuned accelerating structure design will be fabricated in the first year of the demonstration facility,  allowing for engineering refinements of the structure. The accelerating structure can be fully tested outside of the cryomodule,  allowing for multiple iterations during the operation of the demonstration facility if these are needed.

\subsection{A \CCC demonstration facility}


To complete the TDR for the accelerator that we describe here, we will need a full demonstration of the 
\CCC Main Linac technology on the GeV scale.  This demonstration can  be done  in parallel with the preparation of the TDR. The outstanding technical achievements of the ILC, CLIC and NLC collaborations are central to the rapid realization of the \CCC proposal. Many of the subsystems for the accelerator complex are interchangeable between linear collider concepts with manageable modifications to account for variations in pulse format and beam energy.  Because these subsystems are already mature, the \CCC demonstration facility can  focus on the specific  set of technical milestones
associated with the \CCC concept itself:

\begin{itemize}
\item Development of a fully engineered and operational cryomodule including linac supports, alignment, magnets, BPMs,  RF/electrical feedthroughs, liquid and gaseous nitrogen flow, and safety features.
\item Operation of the cryomodule under the full thermal load of the Main Linac and maximum liquid nitrogen flow velocity over the accelerators in the cryomodule, demonstrating an  acceptable level of  vibrations.
\item Operation with a multi-bunch photo-injector to induce wake fields, using 8 high charge bunches 
 and a tunable-delay witness bunch.
\item Achievement of  120~MeV/m accelerating gradient in single bunch mode for an energy gain of 1~GeV in a single cryomodule, including tests at higher gradients to establish breakdown rates.
\item Acceleration and wakefield effect measurements with a fully damped-detuned structure.
\item Development, in partnership with industry, of the baseline  C-band RF source unit that will be installed with the Main Linac. The RF source unit will be modified from existing industrial product lines.

\end{itemize}

A dedicated demonstration facility can complete these milestones within 5 years.  This short timeline is possible because the footprint of the facility is small and can utilize existing accelerator infrastructure.  It can leverage existing DOE test facilities (\eg,  SLAC’s End Station B).  It  would require only 18 commercially-available 50~MW RF sources to achieve the full string test, as described below.   Though we anticipate building additional modules, it should be  possible to finalize the design of the accelerating structure including demonstrations of accelerating gradient and detuning/damping  prior to completion of the cryomodule. 

We plan for a string test of 3 9-meter cryomodules assembled in series.   The 18 RF sources will suffice to power all three cryomodules  at 70~MeV/m, two modules at 120~MeV/m, or one module  at 170~MeV/m.  This will provide a variety of tests of the robustness of the accelerator design.  These include a full structure test at the expected gradient for \CCC 250 and tests of full-power operation at the expected gradient of the \CCC 550, with a minimum of three cryomodules.  Operating the structures at up to 170~MeV/m could provide the margin and breakdown statistics needed to confirm a higher gradient of operation for the energy upgrade to \CCC 550 (for example, 155~MeV/m) in the exact same footprint as \CCC 250 or for a multi-TeV extension of the Main Linac. In any configuration, the facility will test the vibration-tolerance in cryogenic operation.  Modification of the RF source for tunnel-ready mass production can proceed in parallel to the demonstration of the cryomodules. This five-year timeline is well aligned with the timeline for the  preparation of the \CCC TDR,  which would  be informed by the progress of the demonstration facility for the Main Linac specifications. 

Both the demonstration facility and the full TDR would require an O(100M$\$$) investment. For the TDR this is based on the effort required to produce the ILC TDR~\cite{Bambade:2019fyw}. 

This demonstration facility could also be utilized to pursue additional technical milestones which could have significant impact on improving the accelerator complex on the timescale of the TDR. This would include the demonstration of a cryogenic RF photo-injector capable of producing a polarized electron beam with a brightness sufficient to eliminate the electron damping ring, and the development of RF pulse compressors which would simplify cooling requirements (critical for electrical power consumption with expansion of the complex to the multi-TeV scale) and potentially reduce the overall RF source cost.

After completion of the demonstration milestones for \CCC, this facility would continue on with a high impact R\&D program.  It would provide opportunities to further advance the state of the art for \CCC upgrades (improvement of RF sources, testing with different bunch parameters and pulse formats, \etc).  The demonstration accelerator would also provide a source or injector to  enable R\&D opportunities for other advanced accelerator concepts (such as gamma-gamma colliders, plasma wakefield acceleration, structure wakefield acceleration, \etc).  These alternate concepts could also include \CCC technology in their eventual implementation.   The facility would also carry out research on  related accelerator R\&D (FELs, beam dynamics, extreme bunch compression, \etc),  including the possibility of staging multiple cryomodules to achieve higher energy.   

\subsection{Development of an ultra-low-emittance $e^-$ gun}
\label{sec:coldgun}

The significant increase in peak surface fields with cryogenic copper also opens a new frontier in the development of high brightness polarized electron sources~\cite{Rosenzweig2018}. The achieved surface fields in cryogenic copper structures are in principal sufficient to produce an asymmetric-emittance low enough that it would negate the need for the electron damping ring~\cite{Robles2021}. Early stage R\&D for this cryo-RF electron gun is underway, so that  a source of this type could be utilized for the demonstration facility. If fully realized on this timescale, it could be adopted for the baseline TDR \CCC proposal.

\subsection{Low cost RF source development}

As the facility energy increases, RF source cost plays and every larger role. To mitigate this, we can either make more efficient RF accelerator structures (at least until beam loading dominates) or reduce the cost of RF sources. Recognizing this need, the DOE General Accelerator R\&D road-map for RF Accelerators calls out a goal of developing concepts which will be able to reduce the current large-volume unit cost of about \$7.5/peak-kW to \$2/peak-kW of RF. Many concepts are being pursued to achieve this goal \cite{AF7rf}, and the \CCC demonstration facility would be and excellent proving ground for maturing this technology and integrating it in a compatible way with the overall system's RF-only energy upgrade. There would be a 15 year-window from the start of the demonstration facility construction to the energy upgrade in order to develop these concepts and transition them to industry.



\section{\CCC design for 250 and 550 GeV}
\label{sec:500}

In this section, we describe the design of the 250 and 550~GeV colliders in detail. The summary of the Main Linac parameters for \CCC is provided Table~\ref{tab:mainlinacparam}.

\subsection{\CCC Strategy}



Our goal is to deliver the Higgs on the fastest timescale that is practical with an accelerator complex capable of pursuing a rich physics program for decades. With this goal in mind, we envision that the \CCC 250 Main Linac will be built at a gradient of 70~MeV/m. This is approximately the optimal gradient for 7.5~\$/peak-kW RF. This is almost a factor of two lower than the  gradient required for \CCC 550 (120~MeV/m). This lower gradient will provide many benefits:

\begin{itemize}

\item Reduction of  the peak power required to establish the gradient in the structure. This allows for increased beam power at lower energy to boost the luminosity. It reduces the cost of the Main Linac sufficiently to utilize presently available RF source technology.

\item Margins in the main linac design for RF power(8\%) and gradient ($\approx$ 150~MeV/m, with additional RF power) in the first phase of operation.

\item Margins for high availability operation - additional cryomodules (3\%) with complete RF ready for acceleration of the next train.

\item Reduction of commissioning time of the Main Linac.  The breakdown rate for the lower initial gradient has been demonstrated to be exceedingly low. As with most copper accelerators, the achievable accelerating gradient for a fixed breakdown rate increases with operation and conditioning of the accelerator. The 120~MeV/m accelerating gradient at an acceptable breakdown rate\footnote{3x10$^{-7}$/pulse/m} would be achievable well in advance of the energy upgrade.

\item Opportunity for an energy upgrade to proceed in parallel with data collection during the initial Higgs run at 250~GeV. This energy upgrade would leave the cryomodules untouched and would consist only of the addition of RF sources in the Main Linac tunnel. This would virtually eliminate linac construction down time between the 250~GeV run and operation at higher energy.

\item Opportunity for any advances in RF source technology that are achieved during the timeline of the demonstration facility and construction of the accelerator complex to be implemented for the second stage, potentially offering a  dramatic reduction in the cost of the energy upgrade. 

\end{itemize}

The main downside of this approach is the added linac length required, which increases both the tunnel length and the number of cryomodules for \CCC~250. However, the total length of the Main Linac is less than 4~km, and this length would be required later in any event to reach 550~GeV. Because of the relatively low cost of tunnelling and \CCC cryomodules compared to the RF sources presently available, we believe it is preferable to construct \CCC~250 at 70~MeV/m, and then add RF power for a future energy upgrade. 



For the energy upgrade which will occur in parallel to the 250~GeV CM run additional RF source power will be installed on the existing operating Main Linac, and additional magnets will be installed in the beam delivery system (BDS). Additional cryomodules which can provide energy margin for the 250~GeV CM run will be installed either during the final phases of construction or during machine down-time during the operation of the accelerator. The RF power per meter into the structure will increase from 30~MW/m to 80~MW/m during the upgrade with the addition of RF sources and possibly RF pulse compressors increasing the gradient from 70~MeV/m to 120~MeV/m. Utilizing the original margin in length and the additional gradient the new CM energy for \CCC will be 550~GeV.

\CCC is capable of extending its energy reach with additional Main Linac length into the multi-TeV range.  While the specific energy of these future upgrades will certainly be informed by the physics of the Higgs program at 250 and 550~GeV, there is no technical limitation to extending the Main Linac into the multi-TeV range. Operation in a higher energy range ($>$550~GeV) would require a longer, re-optimized BDS.

\begin{table}
\begin{center}
\begin{tabular}{|c | c | c | c | c |  c |} 
 \hline
  Parameter  & Unit &  & & & \\
Note  &  & & & Baseline & Compact \\
 \hline\hline
Center of Mass Energy  & GeV & 91 & 250 & 550 & 550 \\
\hline
Luminosity  & x$10^{34}$ cm$^{-2}$s$^{-1}$  & 0.4 & 1.3 & 2.4 & 2.4\\ 
Single Beam Power  & MW & 0.7 & 2 & 2.5 & 2.5 \\ 
Single Linac Active Length   & km & 0.56 & 1.83 & 2.45 & 1.9  \\ 
Injection Energy Main Linac & GeV &  10 & 10 & 10 & 10 \\
Train Rep. Rate & Hz & 120 & 120 & 120 & 120 \\
Bunch Charge & nC & 1 & 1 & 1 & 1 \\
Flat-Top RF Pulse Length & ns & 700 & 700 & 260 & 195 \\
Bunch Spacing & Periods (ns) &  30 (5.26) & 30 (5.26) & 20 (3.5) & 15 (2.6) \\
Bunches per Train & & 133 & 133 & 75 & 75 \\
Average Current  &$\mu$A & 16 & 16 & 9 & 9 \\
Peak Current & A & 0.19 & 0.19 & 0.3 & 0.385 \\
Loaded Accel. Gradient & MeV/m & 70 & 70 & 120 & 155 \\
RF Power for Structure & MW/m & 30 & 30 & 80 & 140 \\

\hline
 \end{tabular}
\end{center}
\caption{Main Linac parameters for \CCC upgrade path. Luminosity parameters are scaled from ILC/CLIC documentation using beam power, emittance and bunch charge. Two options, \textit{Baseline} and \textit{Compact}, are listed for \CCC~550. The \textit{Baseline} option is the cost optimized operating gradient for $\$$2/peak-kW RF source cost. The \textit{Compact} option selects a higher gradient for a 7~km site. }
 \label{tab:mainlinacparam}
\end{table}

\subsection{Main Linac}

\begin{figure}
\begin{center}
 \includegraphics[trim={0 6cm 0 5.5cm}, clip, width=0.95\hsize]{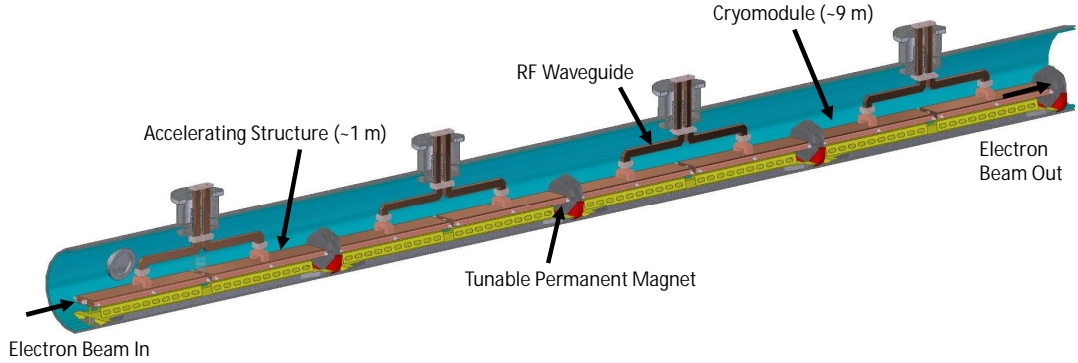} 
 \end{center}
\caption{Cutaway view of a cryomodule.}
\label{fig:cryomodule}
\end{figure}

The Main Linac is constructed from 9~m cryomodules, see Fig.~\ref{fig:cryomodule}, each of which houses eight 1~m distributed coupling accelerating structures supported on four 2~m girders. The girders also support permanent magnet quadrupoles, BPMs, coarse and fine alignment movers, and position sensor fixtures. The cryomodule has four penetrations for RF power, each with two waveguides, each waveguide powering one accelerating structure. The total cryogenic thermal load for the complex at 250~GeV is 9~MW. This thermal load is removed through nucleate boiling of liquid nitrogen.  Liquid nitrogen flows by gravity through the cryomodule to replace the Nitrogen that has evaporated. Thus, the Main Linacs must be horizontal with straight segments of about 1~km. Nitrogen gas is removed from the Main Linac at penetrations that are spaced at approximately 1~km and transported to a re-liquification plant before being reintroduced into the Main Linac.

The goal of the demonstration facility will be to test cryomodules of the final design.


Each cryomodule contains 4 rafts; each raft supports two 1 m accelerator sections and a quadrupole with mechanically integrated Beam Position Monitor. The accelerators and quads are attached to the raft with 5 d.o.f. ball joint adjustable tie rods, and are aligned on the bench before insertion into the cryomodule. The Z position of the midpoint is mechanically fixed.

The rafts have coarse and fine alignment, with each raft having 5 d.o.f. ball joint adjustable tie rods. The Z position of the midpoint is fixed. Each tie rod has both a 200 micron travel piezoelectric actuator, and a magnetically coupled manual screw adjustment from outside the cryomodule. The fine alignment of the rafts will be beam based and dynamic with a bandwidth of $\sim$100 Hz.

The coarse alignment of the rafts is done with stretched wires. Each wire is $\sim$20 m long and spans two cryomodules, with the wires starting on successive cryomodules. Each raft has two wire position sensors for each wire, one on the first accelerator and the other on the quad. The sensors measure X and Y, with a precision of $\sim$100 microns. The rafts are slightly over-constrained with 8 measurements for 5 d.o.f.

The stretched wire method is chosen since it must operate in air, vacuum, and under slowly flowing liquid nitrogen.

\subsection{Main Linac cryogenics}

The increased copper conductivity at $\sim$80~K improves the accelerator efficiency to more than recover the capital and operating expenses of the refrigeration plants. The accelerator is in cryomodules that house 4 rafts; each raft supports 2 accelerator sections and a permanent magnet quadrupole and Beam Position Monitor. 

Sectors consist of 10 cryomodules, super-sectors consist of 10 sectors. (The actual Main Linacs may contain partial super-sectors.)

The cryomodule is a vacuum insulated tube with an inner radius of 30~cm. The accelerator components are under slowly flowing LN at a pressure of $\sim$1.1~bar, with the LN $\sim$25~cm above the cryostat low point. The LN is introduced at the super-sector boundaries or the ends, and flows in both directions (one if at an end) for at most 500 m. Gaseous nitrogen counterflows, is removed at the boundaries or ends, is re-liquified, and then re-injected. The LN for one direction of flow enters at 7.2~l/s, and flows at a velocity (at the entry) of 0.06~m/s (0.2~km/hr).

The LN flow is driven by gravity. The spans between super-sectors are laser straight, the mid-span is normal to the earth radius at its location, and thus the LN is deeper at the center of a super-sector by $\sim$7~cm. The beam will be bent in the vertical plane at the super-sector boundary by $\sim$80~micro-radians to go into the next super-sector. For a 100 GeV beam, this requires a dipole with a kick of 0.05 T-m.

For \CCC~250 and \CCC~550, the power dissipated in one accelerator section is 2500~W, or 0.4$\sim$W/cm$^2$. This is the nucleate boiling regime, and the expected temperature rise is $\sim$2~K. The temperature rise in the copper block in a 1D approximation is 0.6~K.

The counterflow gas flow has an area equivalent to a 23~cm radius pipe, and has a pressure drop over the full length of less than 10~mbar. 

For \CCC~250, the total cooling power for each linac is 4.6 MW, and a cooling run is 508 m. The cooling power per cooling run is $\sim$1.2~MW, which is a medium scale cryo plant. Further optimization is needed to see if fewer but larger plants are more cost effective.

\begin{figure}
\begin{center}
 \includegraphics[trim={0 3cm 0 3cm}, clip, width=0.95\hsize]{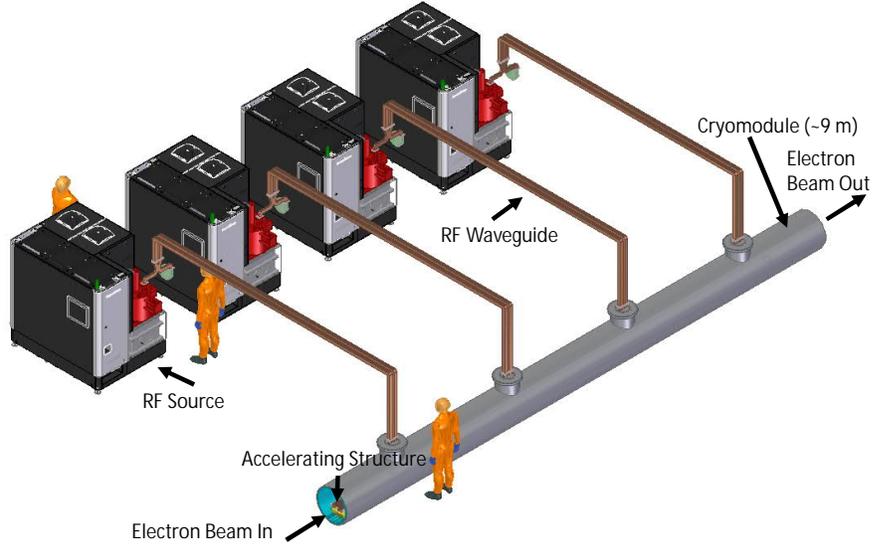} 
 \end{center}
\caption{Cryomodule and four RF sources as installed in the Main Linac.}
\label{fig:cryomoduleandrf}
\end{figure}

For the initial construction of the Main Linac, each of the cryomodules is powered by four RF sources (modulator and klystron), see Fig.~\ref{fig:cryomoduleandrf}. The sources operate at 5.712~GHz and deliver 65~MW for every 2 meters of structure. The power is split in a hybrid and transported to each accelerating structure. The klystron design will be based on advances developed through the High Efficiency International Klystron Activity (HEIKA) collaboration. This includes the implementation of the core oscillation method to retrofit exisitng 50~MW designs to boost power and efficiency \cite{igor2021}. Demonstrations of this design from commercial prototypes at X-band are expected in 2022.

\subsection{Main Linac tunnel layout}
\begin{figure}
\begin{center}
 \includegraphics[trim={0 0cm 0 0cm}, clip, width=0.95\hsize]{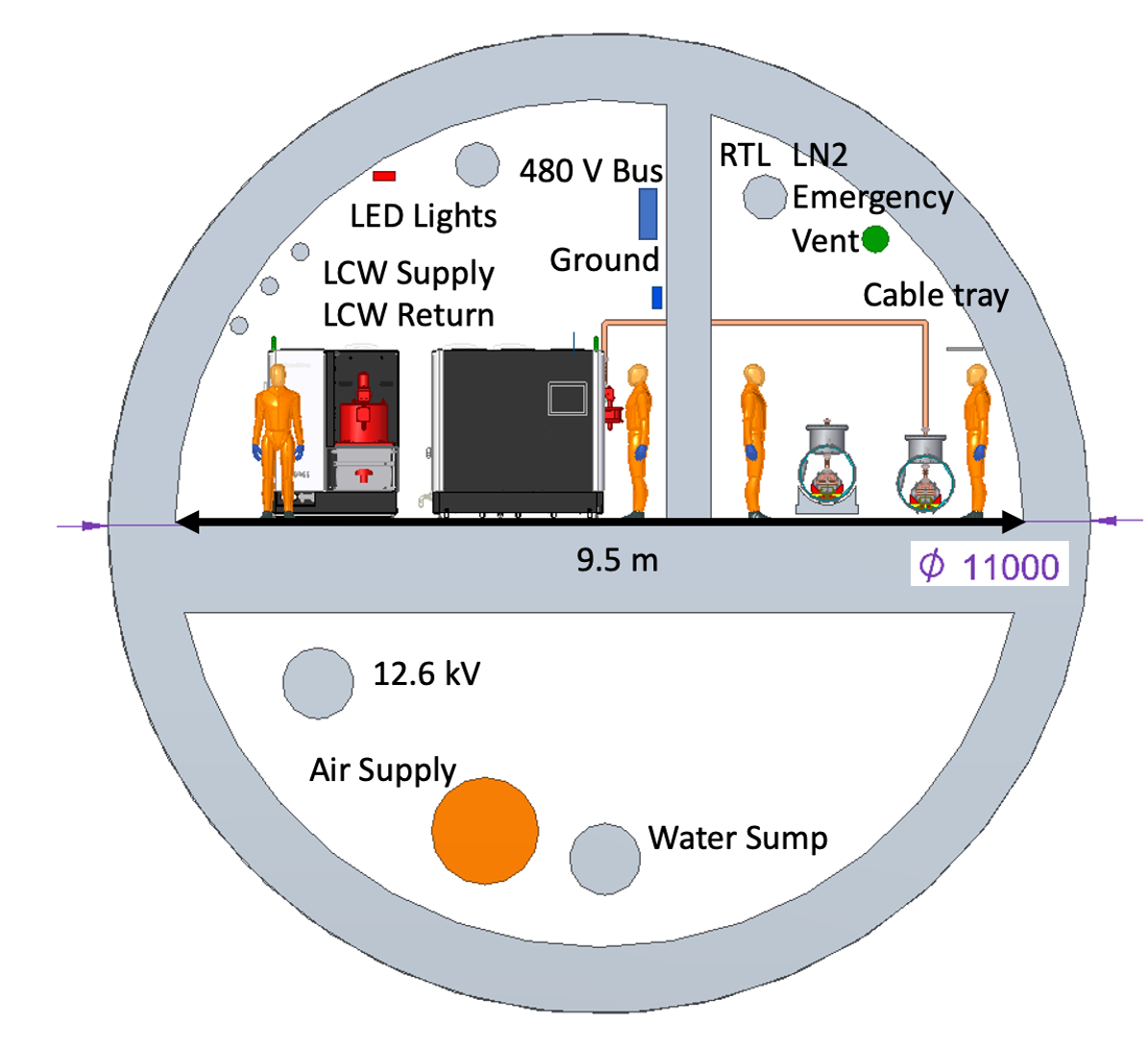} 
 \end{center}
\caption{Cross section of the tunnel layout matching the profile of the ILC tunnel for the combined RF and accelerator gallery. Area shown underneath the gallery floor is not representative of scale. }
\label{fig:tunnellayout}
\end{figure}

The \CCC tunnel layout would be adapted to its location. A cut and cover site, suitable for a horizontal layout, is extremely attractive for both cost and schedule reasons.
The tunnel layout, Fig.~\ref{fig:tunnellayout}, adopts the ILC Asian Region baseline \footnote{ILC TDR Figure 11.6} to match our costing model for the accelerator complex. All major equipment fits in the same cross section as the ILC tunnel. The RF source gallery is separated from the Main Linac by a shielding wall to simplify access. Along with RF sources (klystron, modulator and low level RF), the modulator cooling water and electrical power distribution at 480 V power are in the RF gallery. The accelerator tunnel includes the installed cryomodules, ring to Main Linac transport, possible linac bypass for machine development, and emergency vent lines for the nitrogen in the cryomodules. Air supply, high voltage power distribution and water drainage are housed below the tunnel floor. 
Cross-gallery access is envisioned every 100~m with surface access every 1~km to correspond with penetrations for liquid and gaseous nitrogen. The main linac will also dominate the electrical power consumption for the site. Table~\ref{tab:CCCmainlinacparam} compares these parameters for \CCC 250 and \CCC 550. 

\begin{table}
\begin{center}
\begin{tabular}{|c | c | c | c |} 
 \hline
  Parameter  & Unit & Value  & Value \\
 \hline\hline
Center of Mass Energy  & GeV & 250 & 550 \\
 \hline\hline
Temperature & K  & $\sim$80 & $\sim$80 \\
Pulse Length  & ns & 700 &  250\\
Cryogenic Load $\sim$80~K & MW & 9 &  9\\
Est. RF Power (Both Linacs)  & MW  & 40 &  58\\
Est. Power for Cryogenic Cooling (Both Linacs)    & MW  & 60 &  60\\
Total Est. Power (Both Linacs)    & MW  & 100 &  118 \\
RF Source efficiency (AC line to linac) & $\%$  & 65 & 65\\

\hline
 \end{tabular}
\end{center}
\caption{Main Linac power parameters for \CCC at 250~GeV and 550~GeV center of mass energy.}
 \label{tab:CCCmainlinacparam}
\end{table}

\subsection{Accelerator complex}

\begin{figure}
\begin{center}
 \includegraphics[angle=90,trim={0cm 1cm 0 1cm}, clip, width=0.65\hsize]{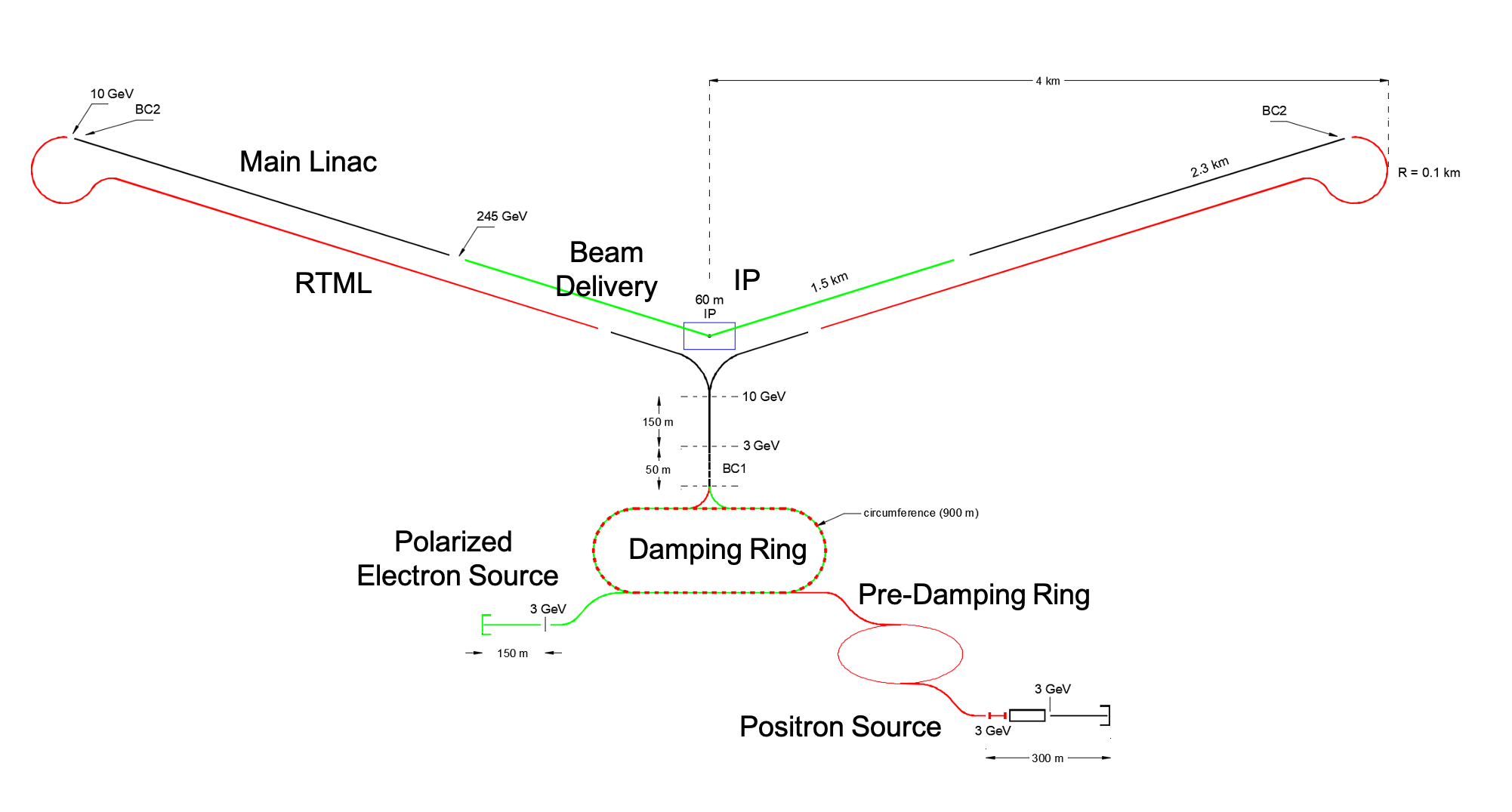} 
 \end{center}
\caption{Conceptual layout (not to scale) for the accelerator complex of \CCC~550. The full accelerator complex fits on an 8~km site.}
\label{fig:AC}
\end{figure}

Great effort has been expended towards the design and optimization of the accelerator complex for earlier linear collider designs. This excellent work can be leveraged to understand the pre-conceptual layout for the full \CCC accelerator complex. However, as the design matures these systems will be re-visited and redesigned for implementation with the \CCC beam format, bunch charge and repetition rate. The electron and positron source, the damping rings, and the BDS are the principal systems of the accelerator complex that are auxiliary to the Main Linac. The conceptual layout of the \CCC accelerator complex is shown in Figure~\ref{fig:AC}.

The baseline electron (polarized) and positron (unpolarized) sources are conventional LC designs. For the electron source, this consists of a polarized DC gun, buncher and accelerator. However, we are also exploring the possibility of a polarized RF gun described in Section~\ref{sec:coldgun}. For the positron source, the closest design relevant to the \CCC bunch structure is the CLIC design\footnote{CLIC CDR 3.1.3.2}. Positron polarization is a possible upgrade once the full RF system is installed. An additional electron bunch train would be extracted from the Main Linac at high energy (125 GeV) and sent to an undulator for polarized photon production and transport to a positron production target. The positron beam must be transported to a damping ring before being accelerated by the Main Linac.

For the damping rings, a conservative design has two damping rings, one for the electrons and one for the positrons. A pre-damping ring is also utilized for the positron beam. Four electron and positron bunch trains are stored in each of the damping rings. The main damping rings have a $\sim$900~m circumference. The electron damping ring might be eliminated with the advent of a polarized RF photoinjector.



\begin{figure}
\begin{center}
 \includegraphics[trim={0 0cm 0 0cm}, clip, width=0.95\hsize]{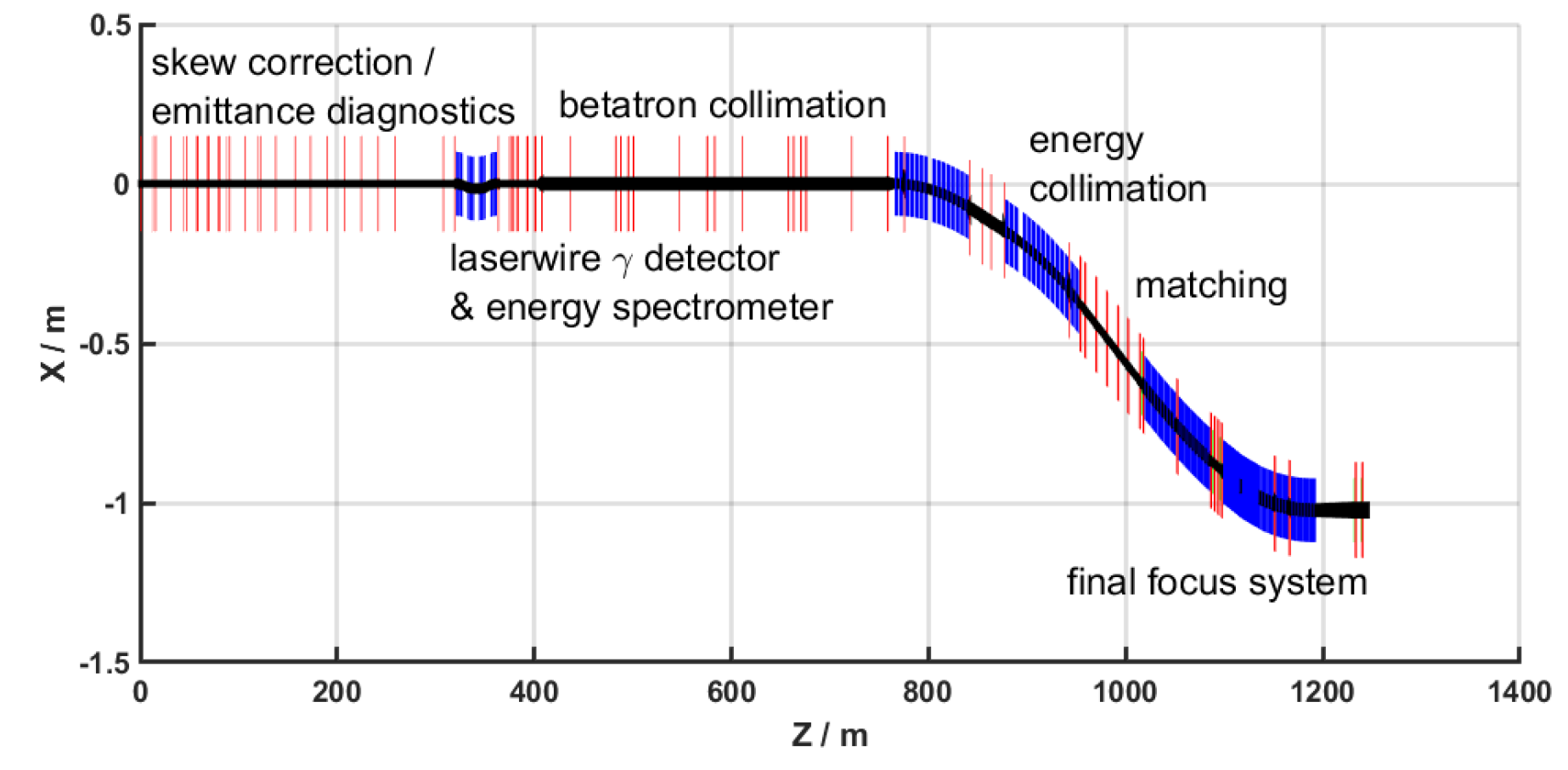} 
 \end{center}
\caption{Preliminary layout of the BDS for 250~GeV CM. An additional 300~m of drift space between components would allow magnets to be added for higher energy operation at 550~GeV.}
\label{fig:BDSlayout}
\end{figure}

\begin{figure}
\begin{center}
 \includegraphics[trim={0 0cm 0 0cm}, clip, width=0.95\hsize]{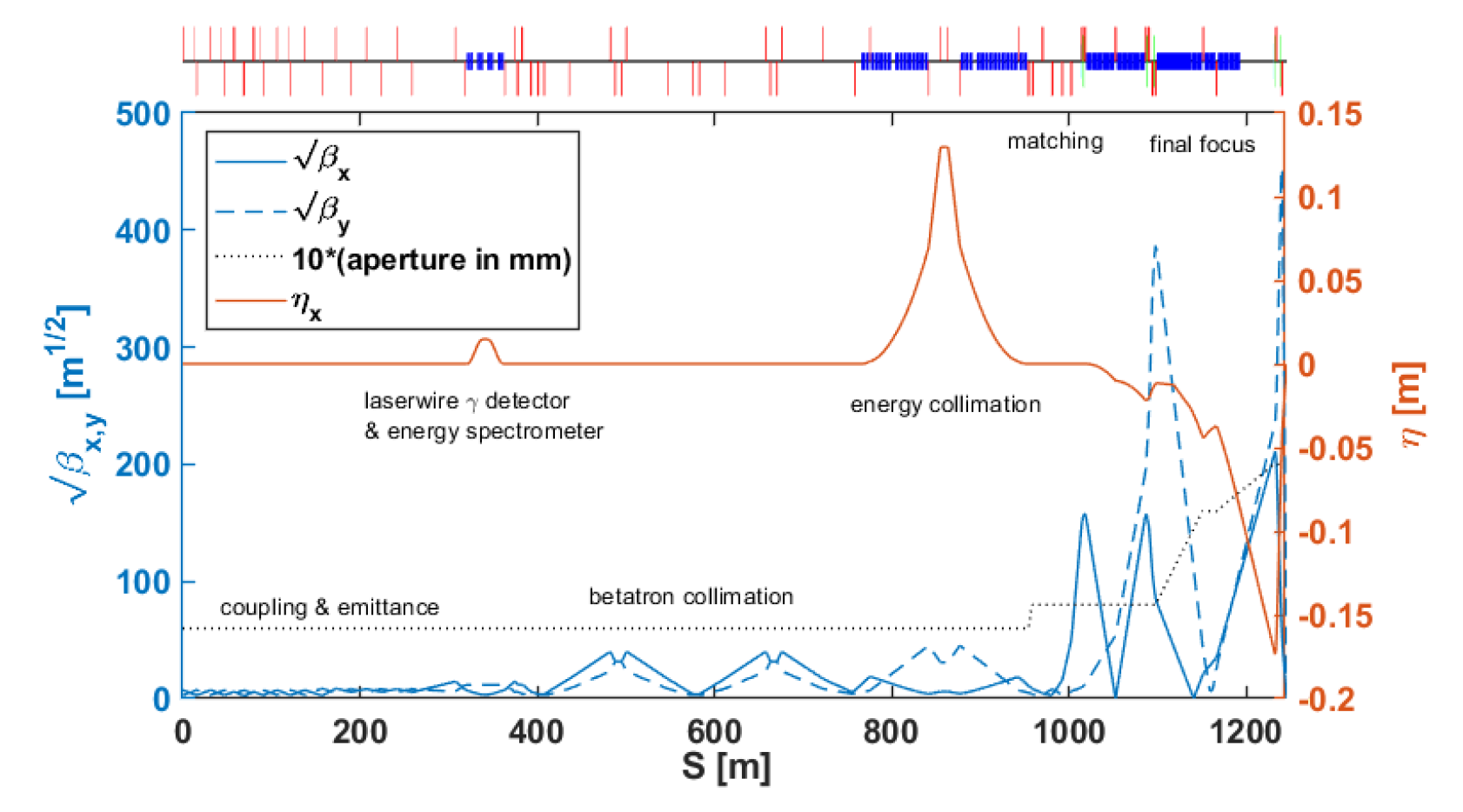} 
 \end{center}
\caption{Optics of the preliminary layout of the 250~GeV CM BDS. }
\label{fig:BDSoptics}
\end{figure}

Scaling of the beam delivery system for a maximum single beam energy of 275~GeV from 500~GeV for the ILC TDR~\cite{Bambade:2019fyw} reduces the length of the BDS by 500~m. Furthermore, we also remove the upstream polarimeter in favor of the downstream polarimeter reducing the length an additional 200~m.  For \CCC, the downstream polarimeter can measure the polarization and the depolarization from beam-beam interactions by comparing interacting and separated beams. The length for the BDS is approximately 1.5~km on each side. Preliminary simulations of for 250~GeV CM indicate that a 1.2~km BDS is feasible, see Figures~\ref{fig:BDSlayout} and \ref{fig:BDSoptics}.

\subsection{Upgrade to 550 GeV}


The completion of the \CCC 250 provides a Main Linac with a full complement of cryomodules capable of operation at 120~MeV/m.  The upgrade to 550~GeV will then mainly entail the addition of RF power sources, to increase the RF power supplied to each meter of linac from 30~MW to 80~MW, doubling the accelerating gradient of the \CCC 250. 

During the construction and commissioning of \CCC~250 we will build a Main Linac tunnel of sufficient length and fully occupied with cryomodules for the upgrade to 550~GeV CM. 

Using present-day cost estimates, the cost of the RF in going directly to  120~MeV/m operation would be 75\% of the of the total linac cost per GeV.  In our staged plan, we achieve savings in two ways, first by constructing the \CCC-250 with RF sources very close to those already produced in large scale by industry, then by relying on the R\&D program described in Sec.~3.4 to provide RF power at much reduced cost for the upgrade.   These additional RF components would be ordered and installed during the \CCC 250 operation, still allowing substantial time to realize the benefits from the new developments.


Additional magnets will be required in the beam delivery system. These magnets will be installed during the machine down time during the physics run for \CCC 250. The length and design of the BDS for \CCC 250 will be sufficient to incorporate the additional magnets for \CCC 550.

\subsection{Cost model}

While developing an accurate cost for any major project is a difficult task, it is particularly challenging in the early stages of a proposed facility. However, given the scale of this endeavor, it is important that we use a cost model to guide our selection of parameters, staging and the overall design of the accelerator complex. In an attempt to provide a cost estimate with reasonable accuracy, we have built a cost model for the accelerator that leverages the ILC and CLIC studies to price key components of the accelerator complex. We rely on ILC for the cost of conventional facilities associated with the subsystems of the accelerator complex (e.g. outfitted tunnel, conventional facilities for damping rings), CLIC for the component costs of the injectors, damping, beam delivery and beam transport and ILC for the IR. 

In addition, we use the cost of fabrication of our prototype \CCC structure to estimate the cost of the cryomodule, and vendor estimates for the cost of the reliquification plants. Similar to ILC and CLIC, here we will quote a Capital Cost for \CCC that includes M\&S, procurement of equipment and construction by qualified vendors. This cost specifically excludes laboratory labor and contingency. The cost is quoted in units ILCU=$\$$=CHF for comparison with other concepts. 
However, we have not completed an estimate of the laboratory labor required for \CCC. This would require a detailed plan with industry vendors given that major components (all RF sources, cryomodules, etc.) could be supplied and installed by external vendors.  

We emphasize that while this exercise was extremely useful in selecting the key parameters and staging plan for \CCC, we believe we have reached the limit of utility for cost scaling and we must now evaluate \CCC specific subsystems of the accelerator complex. This is critical for two reasons. First, the subsystems have some key differences from earlier designs. For example, the injector should utilize the same \CCC technology as the Main Linac.  Also, we have reduced the length of the BDS with a specific energy limit of 550~GeV in order to facilitate fitting the accelerator onto an existing laboratory site. Second, the Main Linac for \CCC 250 GeV CM is no longer the cost driver for the overall accelerator complex. \textbf{This places \CCC 250 in a unique position where cost savings on any subsystem will have an appreciable impact on the overall cost.} We must now revisit the particle sources, BDS and supporting infrastructure to understand how we can simplify and improve to reduce to overall cost of the facility.

Our present estimate for the capital cost of \CCC 250 is in the range of 3.5-4B$\$$ (10\% RF margin, 10 GeV energy margin). The cost breakdown for the accelerator complex is ~35\% sources, ~35\% Main Linac, ~15\% BDS, ~15\% supporting infrastructure. The detailed breakout for one specific scenario is shown in Tab.~\ref{tab:costtable}.

\begin{table}
\begin{center}
\begin{tabular}{|c | c | c | c|} 
 \hline
    & Sub-Domain & \% & \% \\
 \hline\hline
Sources & Injectors  & 8 & 35\\
        & Damping Rings  & 12  &  \\
        & Beam Transport  & 15  &  \\ 
        \hline
Main Linac & Cryomodule  & 10 & 33 \\
           & C-band Klystron  & 23 &  \\
            \hline
BDS & Beam Delivery and Final Focus  & 8 & 13\\
 & IR  & 5 & \\
  \hline
Support Infrastructure  & Civil Engineer &  5 & 19\\
  & Common Facilities &  11 & \\
    & Cryo-plant &  3 &  \\
    
    \hline 
    Total & 3.7B\$ & 100 & 100 \\ 
\hline
 \end{tabular}
\end{center}
 \caption{Cost breakout for \CCC 250 operating at 70~MeV/m. Cost of the outfitted tunnel (51k$\$$/m) and the RF source RF source cost ($\$$7.5/peak-kW), derived from ILC and CLIC respectively, are scaled for the length and RF power needed for the Main Linac. The cryomodule cost of (100k$\$$/m) is based on our production costs.}
\label{tab:costtable}
\end{table}



\section{Extending the \CCC concept to higher energies}
\label{sec:TeV}

We have now described the \CCC concept as a practical and cost-effective way to realize an $\ee$  precision Higgs factory.  But it is also an important aspect of this concept that it offers a path to $\ee$ experiments at much higher energy. 

The precision Higgs program, including the measurement of the top quark Yukawa coupling to 1.6\% and the Higgs self-coupling to 10\%, requires only measurements at CM energies below 1~TeV. In this regime, the klystron-based design that we have proposed here has clear advantages.  The CLIC two-beam accelerator is put forward as a solution for multi-TeV $\ee$ experiments. But \CCC can also be applied here.  The original \CCC paper~\cite{Bane:2018fzj} offered a design of a 2~TeV collider.  In Sec.~\ref{sec:3tev}, we will present a parameter set for a 3~TeV \CCC collider that is a simple extension of our 500~GeV design, using the same gradient and simply making the Main Linac longer.  Either scheme provides a practical way, with known technology, to provide a lepton collider experiment at 3~TeV in the center of mass.  It will be interesting, as these concepts develop, to make a detailed comparison of the two schemes.

Beyond 3~TeV, two additional issues arise for $\ee$ colliders.  First, the beam power needed to obtain the required luminosity grows.  Second, the beam-beam interaction of electron and positron bunches increases, leading to large energy spread and disruption in $\ee$ collisions. However, we see new ideas that can overcome these problems.  We will describe some of these ideas in Secs.~\ref{sec:highergradient} and \ref{sec:beambeam}.   We are optimistic that we can bring these ideas to maturity over the course of the \CCC Higgs program to provide electron linear colliders that advance the technology we have presented here to 10~TeV and above.

There are important physics goals for $\ee$ colliders at energies of order 3~TeV. Probably the most appreciated goal is that of discovering new heavy particles with only electroweak interactions. The mass reach of 1.5~TeV available here goes beyond what is possible at the HL-LHC.  An illustrative benchmark here is the possibility of discovering the Higgsino at a mass of 1~TeV, the mass value at which the pure Higgsino can be thermally produced in the early universe with the relic density observed for dark matter.   The search for this particle is considered to be well beyond the reach of the HL-LHC and is very challenging even for a 100~TeV proton-proton collider~\cite{Low:2014cba}.  A 3~TeV $\ee$ collider can also make precision studies of vector boson scattering at vector boson CM energies above 1~TeV and can extend the measurement of the top quark electroweak form factors to multi-TeV momentum transfers.  Both of these programs give windows into new physics beyond the reach of HL-LHC.  Many more examples are discussed in the CLIC report \cite{deBlas:2018mhx}.  It should be noted that, at 3~TeV, the physics goals of $\ee$ and muon colliders are essentially identical.

A 3~TeV $\ee$ collider will also provide a stepping-stone to electron colliders in the energy region of 10~TeV and higher.   The goal of such colliders will be to discover and study models of electroweak symmetry-breaking at their natural energy scale.  These include models of supersymmetry 
with naturally heavy partner masses such as models with Dirac gauginos and models of Higgs as a Goldstone boson, which require new strong interactions  at the 10~TeV mass scale.  A collider at these energies could also discover leptoquark bosons that might play a role in flavor physics.   Although the beam-beam interaction will be severe in $\ee$ collisions at these energies, a high-energy electron collider can be operated as a $\gamma\gamma$ collider that provides a comparable reach for new particle discovery and polarized beams for the detailed characterization of those new particles.

\subsection{Operation at 3~TeV with 120~MeV/m}
\label{sec:3tev}


Starting from the design parameters of the \CCC 550 collider and simply extending the length, it is straightforward to propose a design for a 3~TeV $\ee$ collider.  As an example, we present a parameter set  assuming the same accelerating gradient, 120~MeV/m.
 Each side of the BDS would need to be extended to 3~km. The main linac length would be 13.5~km on each side. The total footprint of the complex would be 33~km or approximately the length of ILC in the 500~GeV TDR. The design of the main linac would match our previous study at 2~TeV center of mass energy \cite{Bane:2018fzj} and the parameters would scale linearly. Given the timescale of a 3~TeV center of mass energy collider, we assume a reduced RF source cost of $\$$2/peak-kW which is the target cost for the GARD RF Roadmap. 
 
 We emphasize that experimentation at this collider is a direct extension of lower energy experiments. The experimental environment would be close to that already studied in full simulation by the CLIC collaboration~\cite{Roloff:2018dqu}.

\begin{table}
\begin{center}
\begin{tabular}{|c | c | c | c |} 
 \hline
  Parameter  & Unit & Value  & Value \\
 \hline\hline
Center of Mass Energy   & GeV & 250 & 3000 \\
 \hline\hline
Temperature & K  & 77 & 77\\
Main Linac Cost & G$\$$/TeV & 4.8 & 3\\
Accel. Grad. & MeV/m & 70 & 120\\
RF Source Cost & $\$$/peak kW & 7.5 & 2\\
RF Compressors & k$\$$/m  & n/a & 15\\
Pulse Length  & ns & 700 &  250\\
Cryogenic Load at 77 K & MW & 9 &  36.6\\
Est. AC Power for RF Sources  & MW  & 40 &  395\\
Est. Electrical Power for Cryogenic Cooling  & MW  & 60 &  255\\
RF Source efficiency (AC line to linac) & $\%$  & 65 & 65\\

\hline
 \end{tabular}
\end{center}
\caption{Main Linac parameters for \CCC at 3~TeV center of mass energy.}
 \label{tab:TeVmainlinacparam}
\end{table}

\begin{table}
\begin{center}
\begin{tabular}{|c | c | c | c |} 
 \hline
  Parameter  & Unit &  &  \\
 \hline\hline
Center of Mass Energy  & GeV & 550 & 3000 \\
Luminosity  & x$10^{34}$ cm$^{-2}$s$^{-1}$  & 2.4 & 6 \\ 
Single Beam Power  & MW & 2.5 & 13.5 \\ 
Single Linac Active Length   & km  & 2.45 &  13.4 \\ 
Injection Energy Main Linac & GeV  & 10 & 10  \\
Train Rep. Rate & Hz & 120 & 120  \\
Bunch Charge & nC & 1 & 1 \\
Flat-Top RF Pulse Length & ns & 260 & 260 \\
Bunch Spacing & Periods (ns)  & 20 (3.5) &  17 (3.5) \\
Average Current  &$\mu$A  & 9 & 9 \\
Peak Current & A & 0.3 & 0.3 \\
Loaded Accelerating Gradient & MeV/m &  120 & 120\\
RF Power for Structure Flat-Top & MW/m &  80 & 80\\

\hline
 \end{tabular}
\end{center}
\caption{Main Linac parameters for \CCC upgrade path.}
 \label{tab:mainlinacparams2}
\end{table}

\subsection{Ideas for Higher-Gradient Operation}
\label{sec:highergradient}


To reach CM energies of 10~TeV and above, with conventional RF acceleration, we will need to achieve accelerating gradients significantly higher than 120~MeV/m.
The possibility of operating the \CCC accelerator at higher gradients is a tantalizing one. Experiments at both room and low-temperature have significantly exceeded the 120~MeV/m accelerating gradient proposed for \CCC 550. Gradients of up to 250~MeV/m have been achieved in cryogenic RF structures in this frequency range. Unfortunately, the required RF power scales as the accelerating gradient squared which increases the overall cost of the Main Linac. Space limitations, significant reductions in RF power cost, or improved accelerator structure shunt impedance (efficiency) may render the operation at higher gradient more appealing. During the timescale of the demonstration facility we will have the  opportunity to push the accelerating gradient of the \CCC linac well beyond 120~MeV/m, measure the break down rate in realistic operating conditions and make a determination of the physical limit of the installed cryomodules capacity for higher gradient.

Operating beyond 250~MeV/m for an RF accelerator would require a new topology for the linac. One possibility that is actively being researched is the use of shorter RF pulses and higher repetition rates at significantly higher frequencies (100-300 GHz). Structures in this frequency range have exceeded GV/m surface fields. High frequency structures made significant progress in recent years and are now being utilized for beam acceleration and beam manipulation but still require significant R\&D before formulating a proposal for a high energy facility. Extensive investigations into the beam dynamics of such structures are required to confirm the viability for a high luminosity application such as a linear collider. Studies of beam transport in these structures with high gradients (500~MeV/m) and pC bunches indicate it is possible to transport the beam while accounting for effects from short and long range wakes. Long range wakefields are particularly challenging at high frequency. Operating with a bunch in every RF period is one approach to increasing the beam current and allow only for harmonic excitation of higher order modes. The harmonics would need to be at integer multiples of the drive frequency greatly reducing the number of modes which must be damped. With a bunch charge of 1~pC in every cycle, operation at 300 GHz would provide the same peak current during the RF pulse as \CCC 550. Recovering luminosity would require high repetition rates and reduced beam emittance. Powering of such structures would also require very different approach than presently envisioned for \CCC. High frequency RF sources can be extremely efficient with fast-wave cyclotron resonance masers and long pulse formats. Matching these RF sources to high luminosity applications requires the development of active quasi-optical pulse compression which is ongoing. Beam-driven (wakefield) RF sources, as envision for CLIC, may also be a possible option for such high frequency operation.

Finally, although C-band is the optimal frequency for the 250—500~GeV region, it has been discussed since the 1990’s that higher frequencies could be preferred for operation  with smaller bunches to obtain greater power efficiency at higher gradients
 (see, for example, \cite{Zimmermann:1998et}).   Attempts then to realize this approach stumbled on the issue of breakdown.  It is now appealing to revisit this program in the light of the advances from \CCC, to see whether operation at X- or even W-band can give a practical route to very high gradients.

\subsection{Mitigating the $\ee$ beam-beam interaction}
\label{sec:beambeam}

The $\ee$ beam-beam interaction grows rapidly in the multi-TeV region, limiting the luminosity. What can be done about this? First, we are talking here about a discovery, not a precision machine. We can live with 20-30\% spread in the luminosity spectrum.  At  10 TeV energies, the path length 
$c\tau\gamma$ of $b$ and $c$ quarks and $\tau$ leptons is 20 cm or more, so no vertex detector is needed.  The major problem with an $\ee$  design is  then to convey the disrupted beam safely to the beam dump.

To work at the highest luminosities, though, it becomes advantageous to convert 
the electron beam to a photon beam and use the electron collider as a $\gamma$—$\gamma$ collider.
   There are three ways that this conversion can be accomplished.   The first is to  
Compton scattering from the beam of a high-power micron-wavelength laser.   These lasers are undergoing rapid 
development in brightness and repetition rate~\cite{Falcone:2020lou}.    The second is to Compton scattering with a 
Free-Electron Laser as the photon source,  using  a short \CCC accelerator as a driver~\cite{Barklow}.   A third, still rather speculative, 
approach is to use short $\ee$ bunches in the deep quantum regime of beamstrahlung to generate 
the hard photons~\cite{Blankenbecler:1988te,Yakimenko:2018kih}.   There is room here also for many new ideas.

As we pursue the practical goal of building a \CCC Higgs factory, we will also pursue the lines of investigation sketched here toward designing a linear collider for the 10-TeV energy region based on the \CCC concept.  This path starts with many advantages over competing competing technologies.  We are optimistic that it can reach the goal.

\section{Conclusions}
\label{sec:Conclusions}
We present a proposal for a cold copper (cryogenic) distributed coupling linear $\ee$ collider that can provide a rapid route to precision Higgs physics with a compact 8~km footprint.  \CCC is based on recent advances in copper accelerator technology that allow robust designs with an accelerating gradient of 120~MeV/m. \CCC follows a program very similar to that proposed for the ILC, aiming at collecting data at 250 and 550 GeV in the center of mass, with a relatively inexpensive upgrade on the same footprint.  We thus expect to achieve precision measurements of similar quality to those discussed for ILC and other $\ee$ Higgs factories.  This program, starting before 2040, and possibly coincident with the end of the HL-LHC run, will achieve the goals of precision Higgs boson and top quark measurement at which one could realistically prove the presence of deviations from the SM. At the same time, \CCC  will also provide the first step toward the extension of $\ee$ physics into the multi-TeV energy range.

Although there is as yet no engineered and costed design for a 250~GeV $\ee$ \CCC, this approach is based on the SLC experience at SLAC and the extensive design work of ILC and CLIC. Our present estimate for the capital cost of \CCC 250 is in the range of 3.5-4B$\$$ (10\% RF margin, 10 GeV energy margin). The cost breakdown for the accelerator complex are ~35\% sources, ~35\% Main Linac, ~15\% IP, ~15\% supporting infrastructure.  This places \CCC 250 in a  unique position where cost savings on any subsystem will have an appreciable impact on the overall cost.

\Acknowledgements

We are grateful to Fermilab Site Filler Collider working group, Sridhara Dasu, JoAnne Hewett, Tor O. Raubenheimer, Hirohisa A. Tanaka and SiD exectutive committee for their interest and encouragement. We would thank Nan Phinney for the careful revision of the manuscript. The work of the SLAC authors is supported by the US Department of Energy under contract DE–AC02–76SF00515.



\newpage

\end{document}